\documentclass[longauth]{aa}  

\usepackage{graphicx}
\usepackage{color, colortbl}
\usepackage[dvipsnames]{xcolor}
\usepackage{txfonts}
\usepackage[hidelinks=true,colorlinks=true,colorlinks=blue,citecolor=blue]{hyperref}
\usepackage{accessibility}
\definecolor{Gray}{gray}{0.9}

\begin{document}

\title{The ALMA survey to Resolve exoKuiper belt Substructures (ARKS)}

\subtitle{XI: Gas-dust interactions and radial offsets between micron and millimetre-sized grains}

\author{J.~Olofsson\textsuperscript{1}\fnmsep\thanks{E-mail: johan.olofsson@eso.org} \and M.~R.~Jankovic\textsuperscript{2} \and S.~Marino\textsuperscript{3} \and A.~V.~Krivov\textsuperscript{4} \and M.~Bonduelle\textsuperscript{5} \and G.~Cataldi\textsuperscript{6,7} \and Y.~Han\textsuperscript{8} \and A.~M.~Hughes\textsuperscript{9} \and T.~L\"ohne\textsuperscript{4} \and S.~Mac~Manamon\textsuperscript{10} \and E.~Mansell\textsuperscript{9} \and L.~Matr\`a\textsuperscript{10} \and J.~Milli\textsuperscript{5} \and A.~A.~Sefilian\textsuperscript{11} \and P.~Th\'ebault\textsuperscript{12} \and B.~Zawadzki\textsuperscript{9} \and M.~Booth\textsuperscript{13} \and C.~del~Burgo\textsuperscript{14,15} \and J.~M.~Carpenter\textsuperscript{16} \and Th.~Henning\textsuperscript{17} \and J.~B.~Lovell\textsuperscript{18} \and T.~Pearce\textsuperscript{19} \and S.~P\'erez\textsuperscript{20,21,22} \and D.~J.~Wilner\textsuperscript{18} \and M.~C.~Wyatt\textsuperscript{23}} 

\institute{
European Southern Observatory, Karl-Schwarzschild-Strasse 2, 85748 Garching bei M\"unchen, Germany \and
Institute of Physics Belgrade, University of Belgrade, Pregrevica 118, 11080 Belgrade, Serbia \and
Department of Physics and Astronomy, University of Exeter, Stocker Road, Exeter EX4 4QL, UK \and
Astrophysikalisches Institut und Universit\"atssternwarte, Friedrich-Schiller-Universit\"at Jena, Schillerg\"a{\ss}chen 2-3, 07745 Jena, Germany \and
Univ. Grenoble Alpes, CNRS, IPAG, F-38000 Grenoble, France \and
National Astronomical Observatory of Japan, Osawa 2-21-1, Mitaka, Tokyo 181-8588, Japan \and
Department of Astronomy, Graduate School of Science, The University of Tokyo, Tokyo 113-0033, Japan \and
Division of Geological and Planetary Sciences, California Institute of Technology, 1200 E. California Blvd., Pasadena, CA 91125, USA \and
Department of Astronomy, Van Vleck Observatory, Wesleyan University, 96 Foss Hill Dr., Middletown, CT, 06459, USA \and
School of Physics, Trinity College Dublin, the University of Dublin, College Green, Dublin 2, Ireland \and
Department of Astronomy and Steward Observatory, The University of Arizona, 933 North Cherry Ave, Tucson, AZ, 85721, USA \and
LESIA-Observatoire de Paris, UPMC Univ. Paris 06, Univ. Paris-Diderot, France \and
UK Astronomy Technology Centre, Royal Observatory Edinburgh, Blackford Hill, Edinburgh EH9 3HJ, UK \and
Instituto de Astrof\'isica de Canarias, Vía L\'actea S/N, La Laguna, E-38200, Tenerife, Spain \and
Departamento de Astrof\'isica, Universidad de La Laguna, La Laguna, E-38200, Tenerife, Spain \and
Joint ALMA Observatory, Avenida Alonso de C\'ordova 3107, Vitacura 7630355, Santiago, Chile \and
Max-Planck-Insitut f\"ur Astronomie, K\"onigstuhl 17, 69117 Heidelberg, Germany \and
Center for Astrophysics | Harvard \& Smithsonian, 60 Garden St, Cambridge, MA 02138, USA \and
Department of Physics, University of Warwick, Gibbet Hill Road, Coventry CV4 7AL, UK \and
Departamento de Física, Universidad de Santiago de Chile, Av. V\'ictor Jara 3493, Santiago, Chile \and
Millennium Nucleus on Young Exoplanets and their Moons (YEMS), Chile \and
Center for Interdisciplinary Research in Astrophysics Space Exploration (CIRAS), Universidad de Santiago, Chile \and
Institute of Astronomy, University of Cambridge, Madingley Road, Cambridge CB3 0HA, UK
}

  \abstract
   {The dust observed in debris disks is the result of a collisional cascade initiated from $\sim$\,km-sized parent bodies. Using near-infrared to sub-millimeter observations, we can probe particle sizes spanning $2-3$ orders of magnitude, and with sufficient angular resolution we can follow the dynamics of these dust particles. Observations taken as part of the ALMA survey to Resolve exoKuiper belt Substructures (ARKS) program allowed for a detailed comparison with near-infrared scattered light observations, at unprecedented resolution.}
   {The comparison between the two wavelength regimes reveals that for most gas-bearing debris disks, the distribution of small dust grains peaks outward of the distribution of large dust grains. In this paper we investigate whether gas-dust interactions can explain such radial offsets.}
   {We perform numerical simulations accounting for the effects of radiation pressure, gas drag, and collisions, and compute surface brightness profiles at several wavelengths to assess which parameters drive these radial offsets. We explore several families of models, varying the gas mass, disk optical depth, dust size distribution, and radiation pressure strength.}
   {We find that while larger gas masses lead to more efficient outward radial drift, the resulting radial offset strongly depends on the optical depth of the disk, as the drift efficiency directly competes with the particles' collisional lifetime. We also find that increasing the relative number of $\mu$m-sized dust grains usually yields a larger radial offset between scattered light and millimeter observations. Finally, we show that mid-infrared observations can complement near-infrared and sub-millimeter images, and we discuss the formation of secondary rings at near-infrared wavelengths.}
   {The angular resolution achieved by the ARKS program has opened a new avenue to study the dynamics of dust particles in debris disks, revealing unexpected differences between the appearance of the disks scattered light and thermal emission. We showed that gas-dust interactions can explain the observed radial offsets and provide pointers as to which parameters have the most significant impact.}


   \maketitle
%
%________________________________________________________________

\section{Introduction}

A debris disk can be simply described as a collection of large $\sim$\,km-sized bodies and the debris produced by their collisions (\citealp{Krivov2021}). The parent bodies must have formed earlier, in the proto-planetary disk (e.g., \citealp{Lau2024}), and their mutual collisions result in a steep size distribution that can span at least ten orders of magnitude. The small end of the distribution is set by radiation pressure, which quickly removes the smallest particles from the system (\citealp{Krivov2010}). At first glance, describing a debris disk requires just a handful of parameters: a birth ring where the parent bodies are located, a size distribution describing the collisional cascade, and the inventory of forces consisting mostly of gravity and radiation pressure (though more refined models can include stellar multiplicity, the presence of planets, Poynting-Robertson (PR) drag, self-gravity of the disk, or gas drag).

Radiation pressure is a size-dependent force which naturally results in a size segregation as a function of the stellocentric distance $r$. The small dust grains are set on highly eccentric orbits (or can become gravitationally unbound to the central star), forming a halo outside the birth ring, while mm-sized particles are mostly unaffected and remain close to where they were produced. There is therefore a great synergy in comparing observations taken at different wavelengths. Sub-mm ALMA observations trace the large grains and thus the birth ring, while scattered light observations probe the extended halo as well as the birth ring (\citealp{Thebault2023}).
The ALMA survey to resolve exoKuiper belt substructures Large Program (ARKS, \citealp{Marino2026a}) provided new sub-mm observations at high angular resolution, informing about the presence of cold CO gas (\citealp{MacManamon2026}), and enabling a detailed comparison with archival SPHERE images (\citealp{Beuzit2019}). \citet{Milli2026} presented a comparison between radial profiles extracted from the ARKS continuum sub-mm observations (see \citealp{Han2026}) and near-IR scattered light observations. They found that there are six disks with an offset statistically significant and these six disks except one are also known to host cold CO gas. Conversely, the systems without significant offsets are not known to host gas, except one. As discussed in \citet{Takeuchi2001}, \citet{Krivov2009}, \citet{Olofsson2022a}, and more recently \citet{Jankovic2026} and \citet{Weber2026}, gas can alter the dynamics of the dust particles. In this work we therefore investigate whether the presence of gas can explain the presence and strength of the observed offsets. We do not attempt to fit and reproduce any particular set of ALMA and SPHERE observations but rather present a parameter study to check under which circumstances such radial offsets can arise.

In Section\,\ref{sec:unbound} we first describe the simulation setup (from determining the dust density distribution to computing synthetic images). We then present a fiducial model in Section\,\ref{sec:fiducial} and explore the parameter space to quantify which parameters can explain the radial offsets seen in SPHERE and ALMA observations in Section\,\ref{sec:paramspace}. We discuss our findings in Section\,\ref{sec:discussion} before concluding in Section\,\ref{sec:summary}.

%________________________________________________________________

\section{Simulation setup}\label{sec:unbound}

The end goal of each simulation is to compute the dust density distribution as a function of grain size, accounting for the effects of stellar gravity, radiation pressure, gas drag, and collisions. The approach follows the one originally presented in \citet{Thebault2012}, the main differences being the inclusion of gas drag (see also \citealp{Olofsson2022a}) and the fact that unbound grains are included here (similarly to \citealp{Thebault2023}). In this section we first describe the physical parameters defining our model, then explain how the dust density distribution is determined and how synthetic images are computed.

\subsection{Dust and gas parameters}

\subsubsection{Dust parameters}

The full disk can be described using only a few parameters. First, we parametrize the radial distribution of parent bodies in the birth ring by a normal distribution of semi-major axes, centered at $a_\mathrm{d}$ with a standard deviation $\sigma_\mathrm{d}$. Next, the vertical structure is determined by a single parameter $\psi$, the standard deviation of a normal distribution from which we draw the inclinations of the parent bodies with respect to the midplane.

At the beginning of the simulation, $n_\mathrm{part}$ parent bodies are created, which only experience stellar gravity. During the initialization, we draw their mean anomalies from a uniform distribution between $0$ and $2\pi$ and compute their initial positions and velocities. Each of these ``seeds'' parent bodies will release one dust particle (hence $n_\mathrm{part}$ particles in total), whose dynamics we will follow. Furthermore, a radius $s$ is also assigned between a minimum and maximum size, $s_\mathrm{min}$ and $s_\mathrm{max}$, respectively and following a power-law distribution $\mathrm{d}n(s) \propto s^{q}\mathrm{d}s$, with a slope $q$. Finally, each particle is assigned a value of $\beta$, the ratio between radiation pressure and gravity, computed following \citet{Burns1979};
\begin{equation}\label{eqn:beta}
    \beta(s) = \frac{3L_\star}{16 \pi G c M_\star}\frac{Q_\mathrm{pr}(s)}{\rho s},
\end{equation}
where $L_\star$ and $M_\star$ are the stellar luminosity and mass, $G$ the gravitational constant, $c$ the speed of light, $\rho$ the material density of the dust particles, and $Q_\mathrm{pr}$ is the radiation pressure efficiency, computed as 
\begin{equation}
Q_\mathrm{pr}(s) = Q_\mathrm{abs}(s) + Q_\mathrm{sca}(s) \times (1 - g_\mathrm{sca}(s)),
\end{equation}
averaged over the stellar photospheric model, where $Q_\mathrm{abs}$ and $Q_\mathrm{sca}$ are the absorption and scattering efficiencies while $g_\mathrm{sca}$ is the asymmetry parameter. To have finer control over the blow-out size ($s_\mathrm{blow}$ below which grains are unbound to the star), we include another free parameter, $\beta_\mathrm{scale}$, by which we multiply the $\beta(s)$ curve. For $\beta_\mathrm{scale} < 1$, this effectively decreases the blow-out size (shifting down $\beta(s)$) and vice-versa for $\beta_\mathrm{scale} > 1$ (see Fig.\,\ref{fig:beta}). All the quantities related to dust properties are computed using the \texttt{optool}\footnote{Available at \url{https://github.com/cdominik/optool}} package (\citealp{optool}). Unbound grains sent on hyperbolic orbits because of radiation pressure are included in the simulation. 

As will be explained in Section\,\ref{sec:nbody}, the lifetime of the dust particles depends on the optical depth of the disk. This is simply parametrized by a maximum geometric optical depth $\tau_\mathrm{max}$ which will be used to normalize optical depth profiles to estimate the collisional lifetime of the particles. As discussed in \citet{Zuckerman2012}, the optical depth can be related to the fractional luminosity of the disk $f_\mathrm{disk} = L_\mathrm{disk}/L_\star$, as $\tau_\mathrm{max} \sim 2r f_\mathrm{disk}/\Delta r$, where $L_\mathrm{disk}$ is the luminosity of the disk, $r$ is the radius of the disk and $\Delta r$ its radial width.
The last parameters related to the properties of the birth ring are its inclination and position angle ($i$ and $\phi$, respectively). These two parameters relate only to the viewing geometry for computing the images (Section\,\ref{sec:raytracing}), without affecting the physics.

\subsubsection{Gas parameters}

The gaseous disk is parametrized by a normal distribution for the surface density distribution $\Sigma_\mathrm{g}(r)$, centered at $a_\mathrm{g}$ with a standard deviation $\sigma_\mathrm{g}$, and normalized to a total gas mass $M_\mathrm{gas}$. The composition of the gas is described with the mean molecular weight $\mu$, and the gas temperature varies as $T_\mathrm{g}(r) = T_\mathrm{g,0} (r/a_\mathrm{g})^{-1/2}$.

%________________________________________________________________
\subsection{Determining the dust density distribution}\label{sec:nbody}

Since the gas exerts a force on the dust grains which depends on their velocities, we derive the dust density distribution in an iterative manner, as described below.

\subsubsection{Inventory of forces}

At $t=0$, $n_\mathrm{part}$ particles are created and they immediately experience the following three forces: stellar gravity, radiation pressure, and gas drag. The first two are simply accounted for by reducing the stellar mass $M_\star$ by a factor $(1-\beta)$ when computing the acceleration (\citealp{Burns1979}). For the gas drag, we follow \citet{Jankovic2026} and assume that the gas is in vertical hydrostatic equilibrium, with the vertical scale height computed as
\begin{equation}
H_\mathrm{g} = c_\mathrm{s}(r)/\Omega_\mathrm{k} = \sqrt{\frac{k_\mathrm{B} T_\mathrm{g}(r) r^3}{G M_\star \mu m_\mathrm{H}}}, 
\end{equation}
where $c_\mathrm{s}(r)$ is the local sound speed, $\Omega_\mathrm{k}$ the Keplerian angular velocity in the midplane, $k_\mathrm{B}$ the Boltzmann constant, and $m_\mathrm{H}$ the mass of the hydrogen atom. The velocity of the gas in the vertical and radial directions are null, but because the gas feels its own pressure, its azimuthal velocity will depart from the local Keplerian velocity $v_\mathrm{k}$ and can be computed as $v_\mathrm{g} = v_\mathrm{k} \sqrt{1-\eta}$, where $\eta$ is derived as in \citet[][also capped at unity at large separations]{Jankovic2026}:
\begin{equation}\label{eqn:eta}
    \eta(r) = \frac{c_\mathrm{s}^2(r) r}{G M_\star}\left(\frac{7}{4} + \frac{r^2 - r a_\mathrm{g}}{\sigma_\mathrm{g}^2} \right).
\end{equation}
At each distance $r$, there is a value of $\beta (s)$ for which the effect of radiation pressure and gas drag cancel each other out ($\beta (s) = \eta (r)$), so that the particle no longer drifts in the radial direction. This divides the $\beta (s)$ plane in two different domains of inward or outward drift. Figure\,\ref{fig:eta} shows four examples of the $\beta (s)$ stability regions, determined by the condition $\eta(r) = \beta(s)$, for different combinations of $T_\mathrm{g, 0}$ and $\mu$. As discussed in \citet[][see also \citealp{Krivov2009}, \citealp{Jankovic2026}, \citealp{Weber2026}]{Takeuchi2001}, in the absence of collisions, if a dust particle has a $\beta$ value larger than the local value of $\eta$ it will drift outwards, until it reaches a location in the disk where $\beta(s) = \eta(r)$.

The thermal velocity of the gas is computed as in \citet{Takeuchi2001}:
\begin{equation}
    v_\mathrm{therm} = \frac{4}{3}\sqrt{\frac{8 k_\mathrm{B} T_\mathrm{g,0}}{\pi \mu m_\mathrm{H}} \left(\frac{r}{a_\mathrm{g}} \right)^{-1/2}},
\end{equation}
The gas drag force can then be computed as:
\begin{equation}\label{eqn:gdrag}
    F_\mathrm{gas} = -\pi \rho_\mathrm{g} s^2 \sqrt{v_\mathrm{therm}^2 + \Delta v^2}\Delta v,
\end{equation}
where $\rho_\mathrm{g}$ is the local gas density and $\Delta v$ is the relative velocity between the dust particle and the gas, the latter depending on $\eta$.

\subsubsection{Integrating the equation of motion and assessing convergence}\label{sec:convergence}

To follow the motion of the dust particles, we use a fourth order Runge-Kutta integrator and the acceleration accounts for all the aforementioned forces. The timestep $\delta t$ is fixed to one hundredth of the orbital period at a distance of $a_\mathrm{d}$. The final dust density distribution is estimated through an iterative process of several consecutive runs. The purpose of these runs is to make sure that the grain-to-grain collisions are accounted for properly. If one were to compute just one simple iteration, the end result will highly depend on its duration. We therefore follow the approach outlined in \citet{Thebault2012}, which is also quite similar to the one described in \citet{Krivov2009}.

\begin{figure}
  \centering
  \includegraphics[width=1.0\linewidth]{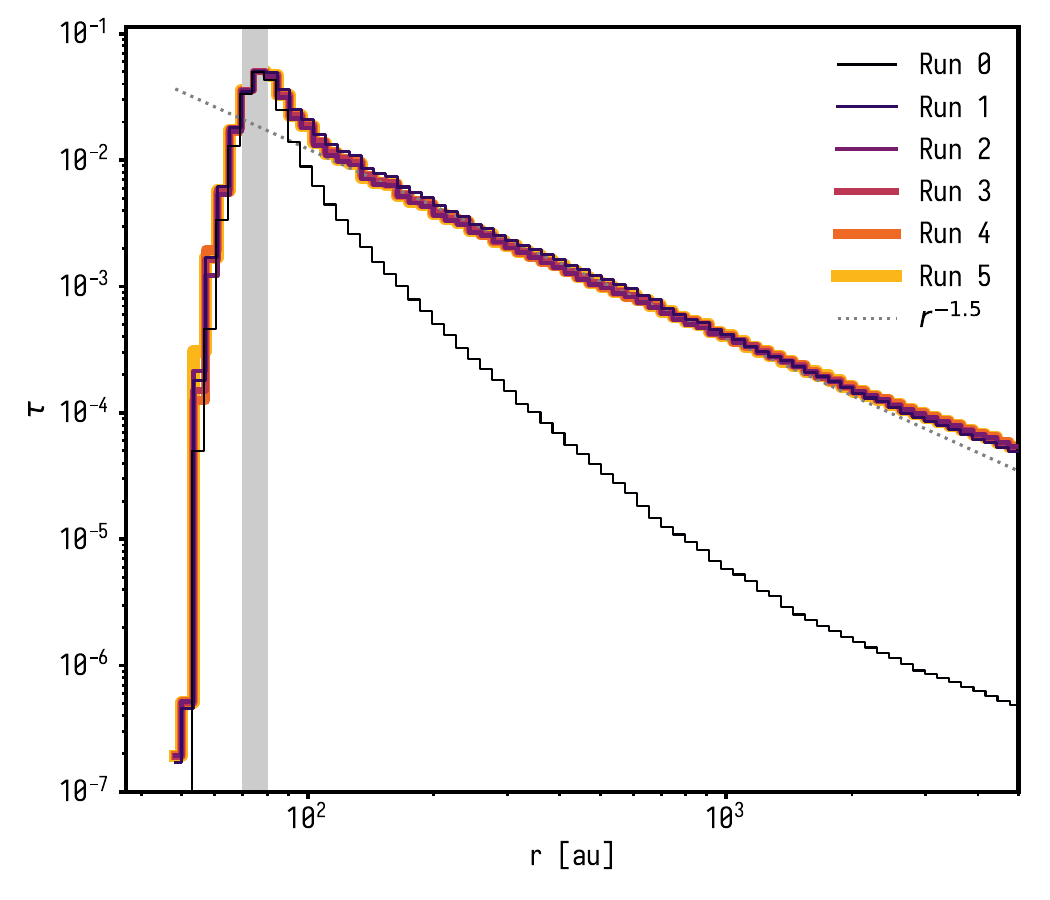}
    \caption{Geometric optical depth as a function of the stellocentric distance. A reference slope in $r^{-1.5}$ is shown with a dotted line. The solid lines of various colors and thicknesses show the evolution of the optical depth for successive iterations (see Section\,\ref{sec:convergence} for details). The location of the birth ring is marked by the grey shaded area ($a_\mathrm{d}\pm \sigma_\mathrm{d}$).
  \label{fig:tau}}
\end{figure}

A first iteration is computed (``run 0'') for an arbitrary duration (e.g., $500\,000$\,years), which will yield a first estimate of the radial optical depth profile. For this first run, there are no collisions and particles are not destroyed. It is parametrized by a total number of integrations and at every timestep we increment the one-dimensional vertical optical depth $\tau_0(r)$ by the cross-section $\pi s^2$ of the particles (meaning it is not computed only at the end of the simulation, but rather updated after each timestep). Once the run is finished, the $\tau_0(r)$ profile is normalized so that the maximum is equal to $\tau_\mathrm{max}$ (an input parameter of the simulation). This is illustrated in Fig\,\ref{fig:tau}, for which $\tau_\mathrm{max} = 5 \times 10^{-2}$. The thin black line shows the estimated optical depth profile at the end of collisionless run. We note that while the code is 3D, since the problem is axisymmetric we use azimuthally averaged quantities whenever possible (e.g., $\tau (r)$) to speed up calculations.

The following iterations ($n$) will use the estimate of $\tau_{n-1}(r)$ of the previous iteration ($n-1$) to determine whether a particle is destroyed or not. For each particle, at each timestep, we compute a probability that a particle is destroyed as
\begin{equation}\label{eqn:tcoll}
    f_\mathrm{coll} = \frac{\pi \delta \mathrm{t} \tau_{n-1}(r)}{t_\mathrm{orbit}},
\end{equation}
similarly to \citet{Olofsson2022a}. If a particle is destroyed, it is removed from the simulation and a collision does not produce new particles. The optical depth profile $\tau_n (r)$ is then updated, at each timestep, with the cross-sections of all remaining particles, and this is repeated until the simulation finishes. A run with collisions will be considered as finished when $99.99$\% of the initial $n_\mathrm{part}$ have been destroyed. The inclusion of unbound particles requires an additional condition for particles to be removed from the simulation. Indeed, these grains are set on hyperbolic orbits and will never come back in the birth ring. Their chances of being destroyed by collisions are therefore very small. We set a maximum distance ($100$ times the reference radius $a_\mathrm{d}$) beyond which we consider the particles to be removed. Overall, the total duration of the simulation for the $n^{\mathrm{th}}$ iteration will depend on the collisional lifetime of the particles, since we need to make sure the vast majority of particles are destroyed one way or another. It may be shorter or longer than in previous iterations, since $\tau (r)$ may differ, especially for the first few iterations. Figure\,\ref{fig:tau} shows the evolution of the optical depth profiles for consecutive runs 0 to 5. It is apparent that the simulation quickly converges to a stable optical depth profile, where runs 3, 4, and 5 are almost perfectly identical to each other. For comparison, a slope in $r^{-1.5}$ is also displayed, which is the expected behavior for the optical depth beyond the birth ring (\citealp{Strubbe2006}, \citealp{Thebault2008}). At large stellocentric distances, beyond $\sim 2\,000$\,au, the profile flattens towards $r^{-1}$ because of the increasing contribution of unbound grains (see \citealp{Strubbe2006}, \citealp{Thebault2023}).

\subsection{Synthetic images}\label{sec:raytracing}

During the last iteration (run\,5), the code will also compute a set of images, in scattered light (total intensity or linear polarimetry) and thermal emission. The images are parametrized by a number of pixels and a pixel size. At each timestep, for each particle, we first apply two rotations to the $(x,y,z)$ coordinates: the first one to account for the inclination $i$ of the disk, and the second for its position angle $\phi$. If the considered particle falls in the field of view of the image, the corresponding pixel is updated as described in the following.

\subsubsection{Scattered light images}

For each particle, we first compute the scattering angle $\theta$ as the arccosine of the dot product between the line of sight and the position vector of the particle. Its contribution is then:
\begin{equation}\label{eq:scattered}
    F_\mathrm{scattered} = \frac{F_\star}{4 \pi r^2} \pi s^2 S_\mathrm{11}(s, \theta) Q_\mathrm{sca}(s)
\end{equation}
where $S_\mathrm{11}$ is the element of the M\"uller matrix for the total intensity phase function, computed with \texttt{optool} (for polarized intensity this would be replaced by $S_\mathrm{12}$), $r$ is the distance to the star, and $F_\star$ the stellar flux.

\subsubsection{Thermal emission}

One key aspect when computing images in thermal emission is to estimate the temperature $T_\mathrm{dust}$ of the particles, which depends on the absorption coefficients (hence $s$ and wavelength $\lambda$) and the distance $r$ to the star. Assuming that the dust grains are at equilibrium temperature, we can equate how much light they absorb and emit, to obtain the following relationship:
\begin{equation}
    r(s, T_\mathrm{dust}) = \frac{R_\star}{2} \sqrt{\frac{\int_{\lambda} F_\star(\lambda) Q_\mathrm{abs}(\lambda , s)\mathrm{d}\lambda}{\int_{\lambda} \pi B(\lambda, T_\mathrm{dust}) Q_\mathrm{abs}(\lambda , s)\mathrm{d}\lambda}},
\end{equation}
where $B$ is the Planck function at the temperature $T_\mathrm{dust}$ and $R_\star$ the stellar radius. When initializing the simulation, the code samples an array of temperatures between $2$ and $3\,000$\,K and computes the corresponding radius for all grain sizes (illustrated in Figure\,\ref{fig:tdust}). During the simulation, when computing the thermal emission image, for a particle of a given size and given distance, we can invert the above equation to find the temperature of the dust grain. If the particle falls in the field of view of the image, the contribution to the flux at a wavelength $\lambda$ then is:
\begin{equation}\label{eq:thermal}
    F_\mathrm{thermal} = 4 \pi s^2 Q_\mathrm{abs}(s, \lambda) \pi B (\lambda, T_\mathrm{dust}).
\end{equation}

\subsubsection{Additional diagnostics}

To facilitate the comparison between scattered light and thermal emission images, we can also save intermediate products, since we have the cross-section and distance for every particle in the simulation. For instance, we save surface brightness profiles as a function of the radius for both kinds of images. This is done by summing the contributions of Equations\,\ref{eq:scattered} or \ref{eq:thermal} in radial bins (with $\Delta r = 0.8$\,au), dividing by the surface area of the considered ring. This is done regardless of the azimuth angles, hence not at a fixed scattering angle in scattered light, a similar approach as de-projecting the image and azimuthally averaging the surface brightness. The contributions to the radial profiles can also be saved for different intervals of $\beta$ so that we can identify which grain sizes contribute most to the final image.

\subsubsection{Convolution, sensitivity, and observables}

For the purpose of this study, neither the images nor the radial profiles are convolved by a point spread function. Since we are mostly interested in the position of the maximum surface brightness, we also opted not to include noise in the final products. While these two aspects could not be overlooked if we were attempting to fit models to the observations, this is not the main goal of this work. Convolution would add further complexity to the multi-dimensional parameter space; for instance, even convolving the scattered light image with a point spread function of fixed width (e.g., $40$\,mas) would have different impacts depending on the distance of the star ($\sim 10$\,pc for a disk like AU\,Mic or $117.9$\,pc for HD\,121617).

Regarding the observables, some of the ARKS papers show radial profiles in surface density (e.g., \citealp{Milli2026}) instead of surface brightness. In this work, we decided that the final products should be as close as possible to the observations and therefore work with surface brightness profiles. 

\subsection{Caveats and known limitations}\label{sec:caveats}

To be able to explore the parameter space, and explain a possible origin of the radial offsets, we need to compute a rather large number of models. To maintain a reasonably short computing time, assumptions and simplifications have to be made. We mention here the main known limitations of the approach.

\subsubsection{Collisional model}\label{sec:caveats_coll}

The collisional lifetime of the dust particles is a simplified estimation. For instance, the \textit{production} of dust grains is assumed to follow a power-law distribution, and all subsequent collisions are purely destructive (the particle will simply be removed from the simulation).

Furthermore, in Equation\,\ref{eqn:tcoll}, the relative cross-sections of the particles do not intervene when computing the probability that a particle is destroyed. In reality, we know that, for a $s^{-3.5}$ size distribution, the global geometrical cross section decreases as $s^{-0.5}$. As a consequence the collisional lifetime should increase as $\sqrt{s}$, thus making larger particles survive longer than smaller ones meaning that our collisional lifetime is under-estimated. Additionally, according to Equation\,\ref{eqn:tcoll}, the relative velocities between two particles are not accounted for when checking whether or not a particle is destroyed. The distribution of relative velocities between dust grains might play a more crucial role in simulations with gas compared to simulations without gas. On the one hand, gas drag should circularize the orbit (decreasing eccentricity while at the same time increasing pericenter distance, see \citealp{Olofsson2022a}). This circularization is expected to lower the relative velocity between two particles since the velocity vectors should become more aligned with each other while the eccentricities of the particles are decreasing. On the other hand, most collisions take place in the birth ring (where the optical depth is maximum), where the velocities of the particles are the largest, regardless of their size. Given the complexity of implementing these latter two effects (cross-sections and velocities) and the underlying assumptions that would go behind such a model, we opt for a simpler solution that still accounts for the dynamical collision lifetime of the particles.

\subsubsection{Gaseous disk}

The prescription for the gaseous disk is also a simplistic description. We assume a normal distribution for the radial distribution of gas. Such a distribution is qualitatively consistent with the observed radial distributions of CO emission in the ARKS sample (\citealp{MacManamon2026}) in the sense that CO density likely decreases both radially inwards and radially outwards from the belt center. We also assume that the gas temperature is the blackbody one and is vertically constant. Furthermore, the gaseous disk  has no velocity in the vertical direction, it does not viscously spread for the full duration of the simulation, and does not contribute gravitationally, despite being assigned a non-zero total mass. \citet{Marino2022} showed that vertical mixing caused by turbulent diffusion could affect the spatial distribution of the dust particles, but since we have very few constraints on the mixing strength we opted not to include such effects in the simulation. Dust radial diffusion due to gas turbulence is not accounted for in the simulations as its strength remains unconstrained.
 
\section{A first model}\label{sec:fiducial}

\begin{figure}
  \centering
  \includegraphics[width=1.0\linewidth]{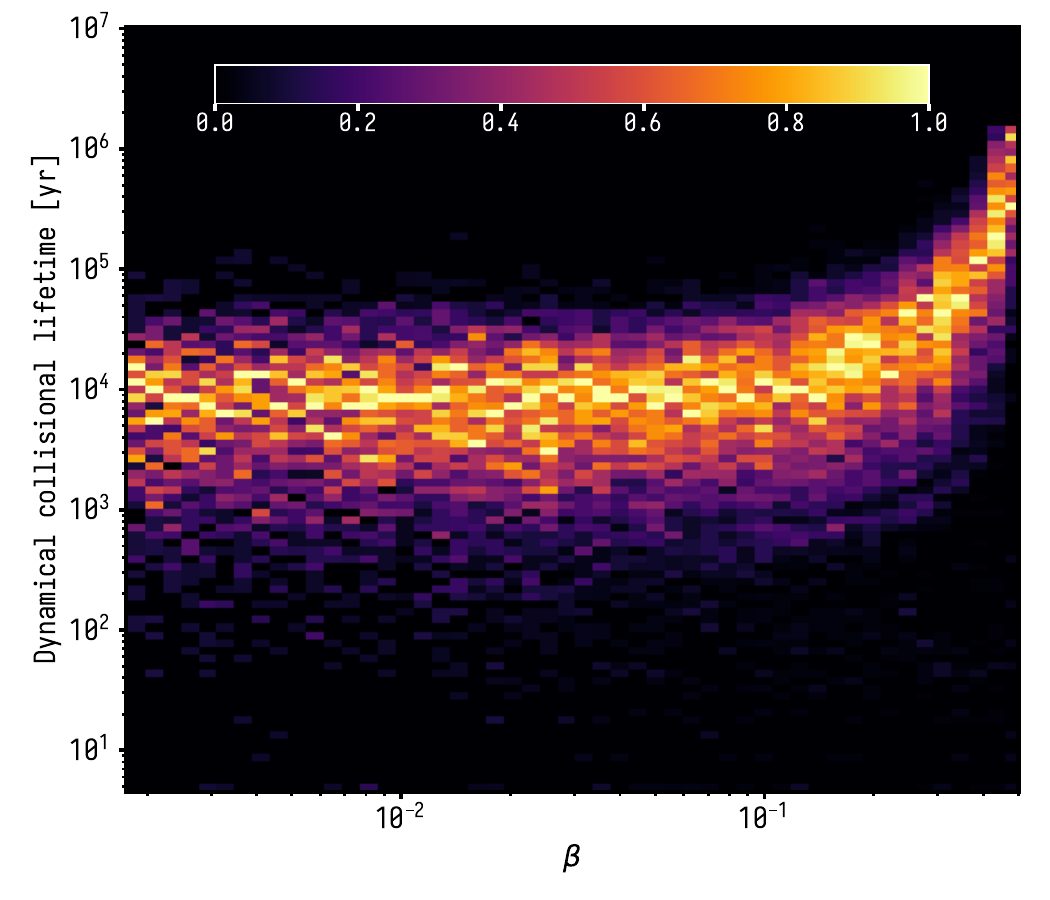}
    \caption{Collisional lifetime as a function of the particles' $\beta$ value. The color-coding indicates the number of particles destroyed in a given cell, re-normalized for each column.
  \label{fig:lifetime}}
\end{figure}

\begin{figure*}
  \centering
  \includegraphics[width=1.0\linewidth]{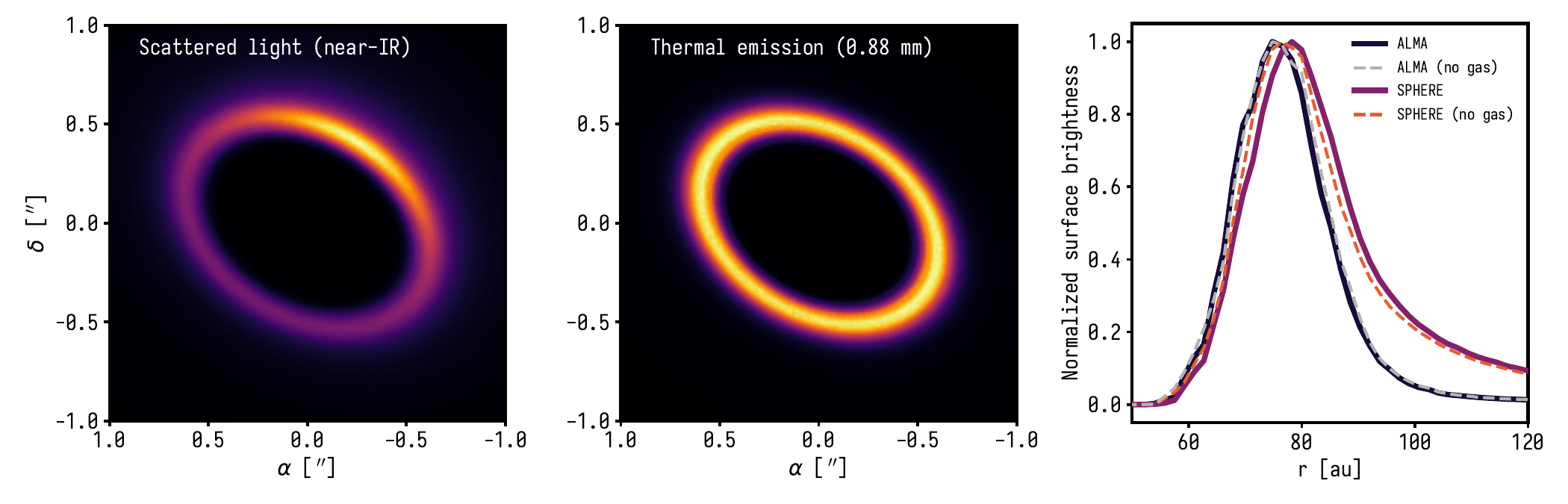}
    \caption{Results for the fiducial model, with a gas mass of $10^{-2}$\,$M_\oplus$. \textit{Left:} synthetic scattered light image in total intensity at $1.63$\,$\mu$m. \textit{Middle:} thermal emission image at $880$\,$\mu$m. \textit{Right:} normalized surface brightness profiles. The profiles for the simulation with gas are shown with solid lines, while the profiles for the gas-free simulation are shown with dashed lines. The ALMA profiles are shown in black and gray, the other two profiles (purple and orange) are for the SPHERE profiles.
  \label{fig:fiducial}}
\end{figure*}

To have ``realistic'' default parameters for a first simulation, we chose HD\,121617, an A-type star with a relatively narrow belt, as a proxy for a typical system hosting a gas-bearing debris disk. Following \citet{Marino2026a}, the stellar parameters are set to $M_\star = 1.9$\,$M_\odot$, $L_\star = 14$\,$L_\odot$, at a distance of $117.89$\,pc. The inclination and position angle of the disk are $i = 44.1^{\circ}$, $\phi = 58.7^{\circ}$. Since we are not aiming to reproduce the ALMA and SPHERE observations of this system, the remaining parameters are defined more loosely. The birth ring of planetesimals is centered at $a_\mathrm{d} = 75$\,au with a standard deviation of $\sigma_\mathrm{d} = 5$\,au. The opening angle of the disk is set to $\psi = 0.04$ and the parent bodies have a proper eccentricity drawn from a normal distribution with a standard deviation of $0.05$ (as in \citealp{Thebault2023} for instance). The peak optical depth is set to $\tau_\mathrm{max} = 5 \times 10^{-3}$, corresponding to $f_\mathrm{disk} = 5\times 10^{-4}$ (assuming $\Delta r = 3 \sigma_\mathrm{d}$). The minimum and maximum grain sizes are $0.1$\,$\mu$m\footnote{\citet{Tazaki2022} showed that aggregates composed of monomers with radii $0.1-0.2$\,$\mu$m can reasonably explain polarimetric observations of the transition disk HD\,142527. We therefore set the minimum grain size to be the size of such monomers.} and $5$\,mm, with a slope in $q = -3.5$, and $\beta_\mathrm{scale} = 1$. The optical properties are computed with the Distribution of Hollow Spheres model (DHS, \citealp{Min2005}) assuming a maximum filling factor $f_\mathrm{max} = 0.8$ and the DIANA composition (\citealp{Woitke2016} and references therein) for which $\rho = 2.08$\,g.cm$^{-3}$. For these parameters, the size for which $\beta = 0.5$ is $\sim 20$\,$\mu$m, meaning that a significant fraction of the size distribution ($0.1 < s < 20$\,$\mu$m) consists of unbound grains. For the gaseous disk, the total mass is $M_\mathrm{gas} = 10^{-2}$\,$M_\oplus$, with a distribution centered at $a_\mathrm{g} = a_\mathrm{d} = 75$\,au, a standard deviation $\sigma_\mathrm{g} = 10$\,au, a mean molecular weight $\mu = 28$ (implying a secondary origin for the gas, where the disk is CO rich and hydrogen poor), and a reference temperature of $T_\mathrm{g, 0} = 40$\,K (\citealp{Brennan2026}).

Figure\,\ref{fig:lifetime} shows the collisional lifetime of particles as a function of the $\beta$ value (only bound grains are considered in this map). When computing the last run of a simulation (run 5 from Fig.\,\ref{fig:tau}), at each timestep, if a particle with a given $\beta$ value is destroyed, we increment the corresponding $(t, \beta)$ cell by unity. The color-coding in Fig.\,\ref{fig:lifetime} therefore shows the number of collisions at a given $(t, \beta)$, re-normalized to the maximum for each $\beta$ value. In this first simulation, most of the particles with $\beta \leq 0.1$ will only survive for about $10,000$\,years. As mentioned in Section\,\ref{sec:caveats_coll}, the probability of having a destructive collision does not depend on the cross-section of the dust particles (Eqn.\,\ref{eqn:tcoll}). The combined effect of both radiation pressure and gas drag are the only processes that could lead to a size-dependent collisional lifetimes. Since particles with low $\beta$ values are less sensitive to these forces, this yields a constant collisional lifetime. But, as $\beta$ increases, so does the lifetime of the particles, since they spend a significant amount of time outside the birth ring, where densities are lower. 

For this first model, we compute two images, one in scattered light, total intensity, at a wavelength of $1.63$\,$\mu$m and a second one in thermal emission at $880$\,$\mu$m\footnote{For brevity we will refer to the two different kind of images as the ``SPHERE'' and the ``ALMA'' images. Furthermore, for SPHERE we will only look at the total intensity images, not polarized intensity.}. As a benchmark, we also compute another simulation where the gas mass is set to $0$. Figure\,\ref{fig:fiducial} shows the results for the set of parameters described above, for the simulation with gas. The left and middle panels show the images in scattered light and in thermal emission while the rightmost panel shows different surface brightness profiles. 

Figure\,\ref{fig:fiducial} shows that the radial profiles for the simulation with gas peak at the following distances: $78.5$\,au for SPHERE, and $75.5$\,au for ALMA ($3.0$\,au offset). The results from the gas-free simulation show that the ALMA profile remains largely unaffected by gas drag, for this set of parameters. The SPHERE profile of the simulation with gas peaks $\sim 1.7$\% further out relative to the simulation without gas ($77.2$\,au), indicating that gas-dust interactions are shifting the distribution of small grains outward.

\begin{figure}
  \centering
  \includegraphics[width=1.0\linewidth]{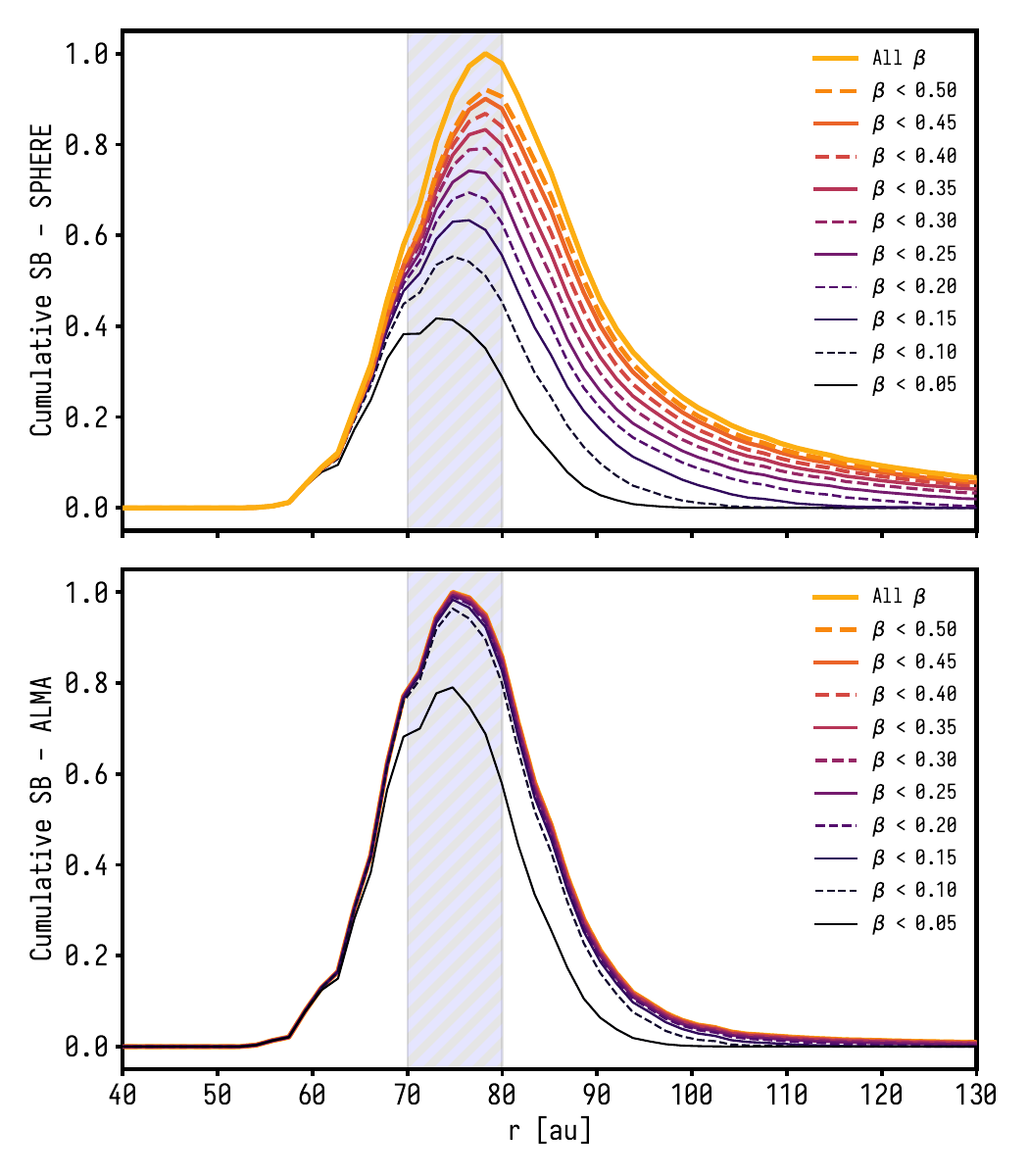}
    \caption{Cumulative contributions to the surface brightness radial profiles as a function of $\beta$ for SPHERE (\textit{top}) and ALMA (\textit{bottom}), for the fiducial model, with $M_\mathrm{gas} = 10^{-2}$\,$M_\oplus$. The hatched area corresponds to $a_\mathrm{d} \pm \sigma_\mathrm{d}$.
  \label{fig:cumulative}}
\end{figure}

Figure\,\ref{fig:cumulative} shows the cumulative contributions to the surface brightness as a function of the $\beta$ value, so that we can investigate which grain sizes are contributing to these radial profiles. This is a cumulative plot in the sense that the line $\beta < 0.15$ also includes the contribution of grains with $\beta < 0.1$ and so on. The lower panel of Fig.\,\ref{fig:cumulative} is for the ALMA image and shows that the majority of the flux comes from grains with $\beta < 0.1$ (i.e., $s > 100$\,$\mu$m) with a marginal contribution from grains smaller than that. In scattered light, the top panel of Fig.\,\ref{fig:cumulative}, the situation is quite different as all intervals contribute to the image. As $\beta$ increases, the peak of the profile shifts to larger distances due to the combined effect of radiation pressure and gas drag. Interestingly, even though they are short-lived, unbound grains do contribute a non-negligible fraction to the image (they make up for the difference between the lines marked "All $\beta$" and "$\beta < 0.5$"), in line with the findings of \citet{Thebault2019}.

\section{Parameter space exploration}\label{sec:paramspace}

\begin{table*}
\centering
    \caption{Summary of the different families of models explored in this study.}
\begin{tabular}{ccccccccccc}
\hline
Family & Number of & $M_{\rm gas}$ & $\tau_{\rm max}$ & $q$ & $\beta_{\rm scale}$ & $a_\mathrm{d}$ & $a_\mathrm{g}$ & $\mu$ & $T_0$  & $s_{\rm min}$\\
       & models    & $[\times 10^{-3} M_\oplus]$ & $[\times 10^{-3}]$ & & & [au] & [au] & & [K] & [$\mu$m]\\
\hline
F1 & 35 & [1, 5, {\color{Orange} 10}, 50, 100, 500, 1000] & [0.5, 1, {\color{Orange} 5}, 10, 50] & -3.5 & 1 & 75 & 75 & 28 & 40 & 0.1 \\
\rowcolor{Gray} F2 & 35 & [1, 5, 10, 50, 100, 500, 1000] & [0.5, 1, 5, 10, 50] & -3.5 & \textbf{0.25} & 75 & 75 & 28 & 40 & 0.1 \\
F3 & 35 & [1, 5, 10, 50, 100, 500, 1000] & [0.5, 1, 5, 10, 50] & \textbf{-3.75} & 1 & 75 & 75 & 28 & 40 & 0.1 \\
\rowcolor{Gray} F4 & 35 & [1, 5, 10, 50, 100, 500, 1000] & [0.5, 1, 5, 10, 50] & \textbf{-3.75} & \textbf{0.25} & 75 & 75 & 28 & 40 & 0.1 \\
F5 & 35 & [1, 5, 10, 50, 100, 500, 1000] & [0.5, 1, 5, 10, 50] & \textbf{-3.75} & \textbf{0.25} & 75 & 75 & 28 & 40 & \textbf{0.5} \\
\rowcolor{Gray} F6 & 35 & [1, 5, 10, 50, 100, 500, 1000] & [0.5, 1, 5, 10, 50] & \textbf{-3.75} & \textbf{0.25} & 75 & 75 & 28 & \textbf{300} & 0.1 \\
F7 & 35 & [1, 5, 10, 50, 100, 500, 1000] & [0.5, 1, 5, 10, 50] & \textbf{-3.75} & \textbf{0.25} & 75 & 75 & \textbf{2} & 40 & 0.1 \\
\rowcolor{Gray} F8 & 35 & [1, 5, 10, 50, 100, 500, 1000] & [0.5, 1, 5, 10, 50] & \textbf{-3.75} & \textbf{0.25} & 75 & \textbf{70} & \textbf{2} & 40 & 0.1 \\
F9 & 35 & [1, 5, 10, 50, 100, 500, 1000] & [0.5, 1, 5, 10, 50] & \textbf{-3.75} & \textbf{0.25} & \textbf{50} & \textbf{50} & 28 & 40 & 0.1 \\
\hline
\end{tabular}
    \tablefoot{For each family the parameter(s) that differ from the fiducial model (which belongs to family F1, highlighted in orange) are highlighted in bold font. The second column reports the number of models computed for each family.}
\label{tab:model_families}
\end{table*}

Even though our approach has relatively few free parameters, computing one simulation can be lengthy (between a few hours to a couple of days). We therefore cannot compute a fully populated grid of models and can only explore combinations of a few parameters. We here follow an empirical approach to identify which ones drive the extent of the radial offset, guided by which ones can either affect the dynamics of the grains, or the integrated optical properties. Table\,\ref{tab:model_families} summarizes the different families of models we compute, and the parameters that are varied with respect to the fiducial model are highlighted in bold font.

\subsection{Gas mass and dust optical depth}

\begin{figure}
  \centering
  \includegraphics[width=1\linewidth]{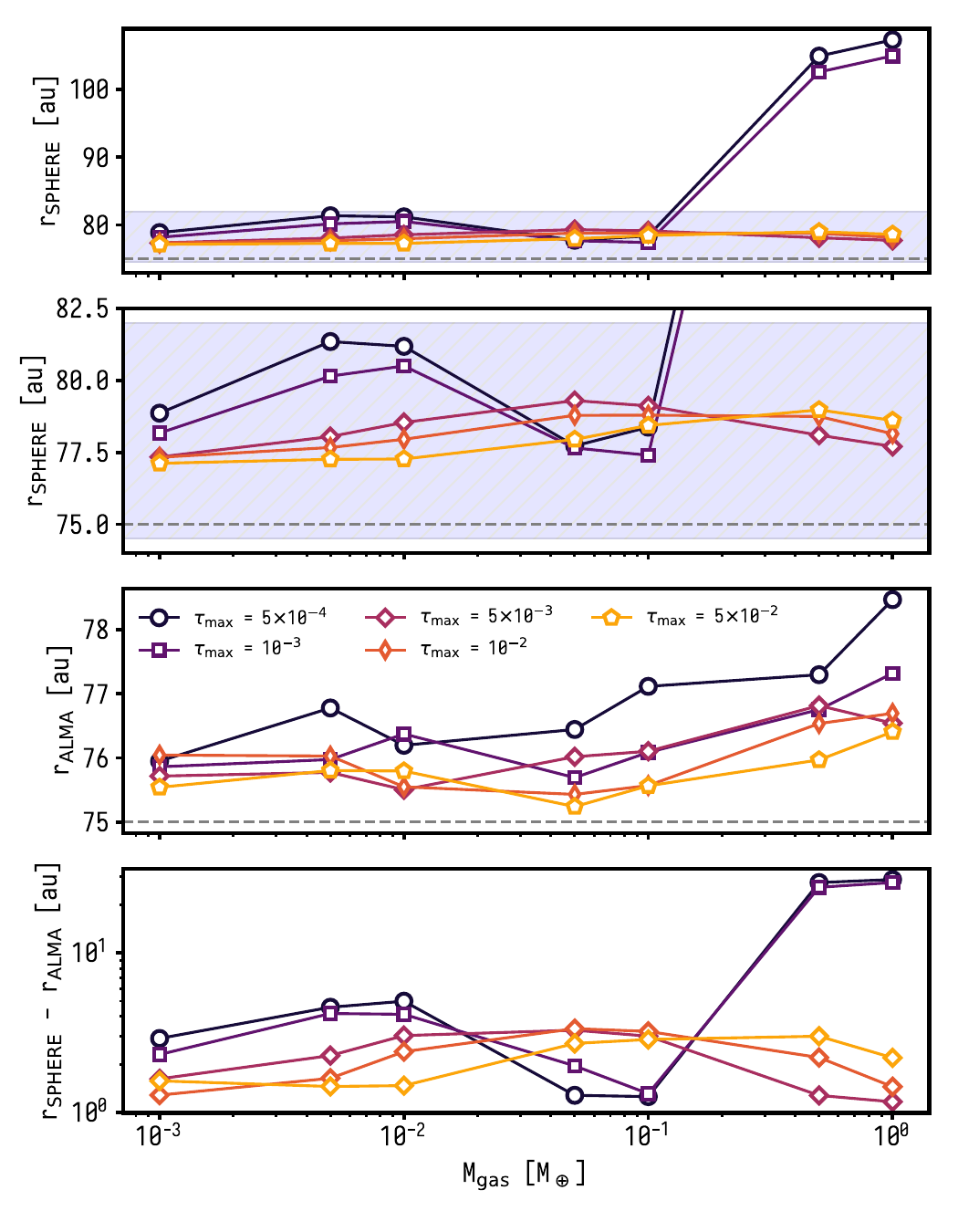}
    \caption{Peak positions for the radial profiles for SPHERE (first and second from the \textit{top}) and ALMA (third from the \textit{top}) and difference between the two (\textit{bottom}, in log-scale), as a function of the total gas mass and for different values of the peak optical depth (legend in the middle panel), for family F1. The second panel from the top shows a zoomed in version of the topmost panel. For the top three panels the horizontal dashed lines show $a_\mathrm{d} = 75$\,au.
  \label{fig:offset_mgas_tau}}
\end{figure}

The first parameters that we explore to vary the radial offset between the SPHERE and ALMA images are the gas mass $M_\mathrm{gas}$ and peak optical depth $\tau_\mathrm{max}$ of the dust, since they will affect either the radial drift efficiency or the collisional timescale. Starting from the first model described in Section\,\ref{sec:fiducial}, we change the gas mass $M_\mathrm{gas}$ to be between $10^{-3}$ and $1$\,$M_\oplus$ (7 values). We also vary $\tau_\mathrm{max}$ between $5 \times 10^{-4}$ and $5 \times 10^{-2}$ (5 values), resulting in $35$ models (the other parameters remaining the same as before, see family F1 in Table\,\ref{tab:model_families}). For this set of parameters, the fractional luminosities of the disks would range between $5 \times 10^{-5}$ and $5 \times 10^{-3}$. 

The peak positions and the offsets for these $35$ models are displayed in Figure\,\ref{fig:offset_mgas_tau}. From top to bottom, we show the peak positions of the SPHERE and ALMA profiles (for clarity, the second panel shows a zoomed view of the top one), as well as the difference between the two, as a function of the gas mass. For each panel, the five curves are for the different values of $\tau_\mathrm{max}$.
For the low gas mass models, the peak positions of the SPHERE and ALMA profiles lie close to the reference radius $a_\mathrm{d} = 75$\,au. For both wavelengths, as the gas mass increases the peaks shift to larger stellocentric distances. We note that for the ALMA profiles (third panel) there is some dispersion on the peak positions ($\lesssim 0.5$\,au). This is because the ALMA surface brightness profiles mostly trace the largest grains (Fig.\,\ref{fig:cumulative}), which are less numerous in our simulations given the steep size distribution, leading to larger numerical Poisson noise for the ALMA profiles compared to the SPHERE profiles. 

As shown in the lower panel of Figure\,\ref{fig:offset_mgas_tau}, the offset is always positive ($r_\mathrm{SPHERE} \geq r_\mathrm{ALMA}$). It first increases with the gas mass but tends to decrease as $M_\mathrm{gas}$ becomes larger, as even the large grains start to drift outward. For the larger gas masses, the offset increases significantly, reaching values close to $30$\,au. This happens when a ``secondary ring'' begins to appear in scattered light. Such secondary rings are known to arise from gas-dust interactions (\citealp{Takeuchi2001}, \citealp{Jankovic2026}). In brief, as discussed in \citet{Jankovic2026}, the drift velocity of the dust particles, under the effects of stellar gravity, radiation pressure, gas drag, and Poynting-Robertson drag, strongly depends on their sizes. The location of the secondary ring is mostly determined by where the drift velocity of the particles falls to zero, which depends on gas pressure profile and density, which in turn depends on gas mass. In this work, we also account for collisions, and the cross-sections of the dust grains will drive the shape of the optical depth radial profile, which will in turn govern the collisional lifetime of the particles, and hence the location of any secondary ring that may form, meaning that the location of the secondary ring also depends on $\tau_\mathrm{max}$. Since a simulation is considered as finished when the optical depth profile $\tau(r)$ no longer changes between consecutive iterations, the secondary ring will not continue to evolve over time. Its brigtness and detectability will however strongly depend on the cross-sections of the grains, their scattering efficiencies (for scattered light images), and their absorption efficiencies and temperatures (for thermal emission images). Since this is investigated in depth in \citet{Jankovic2026}, in the rest of this paper our main focus lies on disks with a single ring, but we still discuss our simulations showing secondary rings and the differences of our approach with the one of \citet{Jankovic2026} in Section\,\ref{sec:HD131835}. 

Increasing gas mass and optical depth seems to have an opposite effect on the peak position of the radial profiles, which can be seen in the third panel of Figure\,\ref{fig:offset_mgas_tau}. The curve for models with high optical depth ($\tau_\mathrm{max} = 5 \times 10^{-2}$ in yellow) is below the one for models with low optical depth ($\tau_\mathrm{max} = 5 \times 10^{-4}$ in black). Regardless of the gas mass, the peak positions of the ALMA profiles peak farther out for low optical depth compared to larger optical depth; the dust particles have more time to drift before they are eventually destroyed by collisions. Furthermore, on the lower panel of Fig.\,\ref{fig:offset_mgas_tau}, the curves for simulations with $\tau_\mathrm{max} = 5 \times 10^{-4}$ and $10^{-3}$ show a maximum offset for a gas mass of $\sim 10^{-2}$\,$M_\oplus$, and the offset decreases for larger gas masses (barring the last two points). As $\tau_\mathrm{max}$ increases, the maximum offset is reached for larger gas masses, closer to $0.1$\,$M_\oplus$, the offset then decreasing for larger gas masses. This correlation between the two parameters is further discussed in Section\,\ref{sec:gmas_tau}.

\subsection{The size distribution}\label{sec:size}

\begin{figure}
  \centering
  \includegraphics[width=1\linewidth]{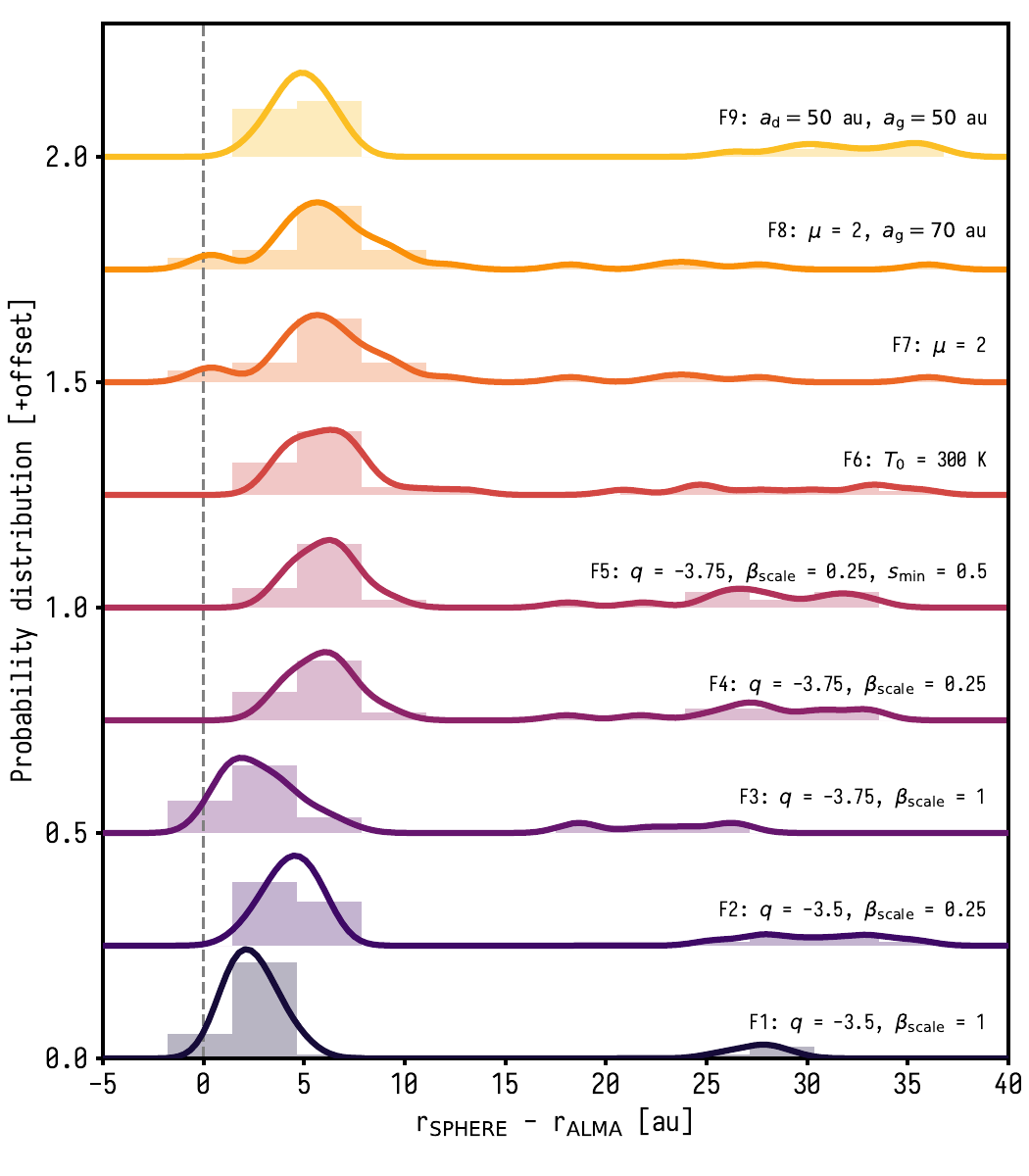}
    \caption{Kernel density estimation and histograms of the offsets for all families of models (F1 to F9). The vertical dashed line is centered at $0$. The fact that some of the curves display negative offsets is due to the fixed kernel's width of $1$\,au.
  \label{fig:kde_size}}
\end{figure}

The surface brightness in the SPHERE images strongly depends on the detailed dust properties, especially for the small end of the size distribution (Fig.\,\ref{fig:cumulative}). Since this would increase the parameter space significantly, we opt not to change the composition or porosity of the dust particles. To test if we can obtain larger offsets between SPHERE and ALMA, we vary parameters that are related to the particle size distribution, which will affect the $Q_\mathrm{sca}$, $S_{11}$, and the cross-sections $s^2$. The first two parameters that we consider here are the slope of the size distribution $q$ and the strength of the radiation pressure via $\beta_\mathrm{scale}$.

First, changing the slope of the size distribution $q$ (to values smaller than $-3.5$) will increase the number of small particles in the simulation. Such steeper values are indeed expected when taking into account the size-dependence of the crushing energy (\citealp{Gaspar2012}). Since the small particles are the ones most strongly affected by radiation pressure and gas drag, increasing their numbers relative to larger grains should affect the peak position of the scattered light surface brightness profile. Second, as mentioned before, the motivation to introduce the $\beta_\mathrm{scale}$ parameter is that it allows us to have a finer control on the $\beta(s)$ relationship. From Eqn.\,\ref{eqn:beta}, this means that $\beta (s)$ is no longer self-consistent with the dust properties $Q_\mathrm{abs}$ or $Q_\mathrm{sca}$ (or the stellar parameters $L_\star$ or $M_\star$). In principle, changing $\beta (s)$ to have a different value for $s_\mathrm{blow}$ can be achieved by changing either the dust composition or the porosity (e.g., \citealp{Arnold2019}), but since the $Q_\mathrm{sca}$ and $S_{11}$ strongly depend on the composition, this would not allow us to isolate the impact that $s_\mathrm{blow}$ has on the final images. Instead, by setting $\beta_\mathrm{scale}$ to \textonequarter, the blow-out size decreases from $20$\,$\mu$m to $5$\,$\mu$m (F2 and F4), while all the other optical properties remain the same. This break of consistency with respect to Eqn.\,\ref{eqn:beta} was deemed acceptable to limit the number of free parameters.

There is a third parameter, related to the size distribution, that we can change; the minimum grain size $s_\mathrm{min}$ (usually $s_\mathrm{min} < s_\mathrm{blow}$ and all the grains with sizes in between would be gravitationally unbound), and family F5 is based on the parameters of F4, only with $s_\mathrm{min}$ changed from $0.1$\,$\mu$m to $0.5$\,$\mu$m.

We consider five families of models (F1 to F5). For each family, we vary $M_\mathrm{gas}$ and $\tau_\mathrm{max}$, as described in Table\,\ref{tab:model_families}, but with $q = -3.5$ or $-3.75$ (close to the slope reported in \citealp{Gaspar2012}), $\beta_\mathrm{scale} = 1$ or \textonequarter, and with $s_\mathrm{min} = 0.1$ or $0.5$\,$\mu$m. For each model we compute the offset between the peak positions of the SPHERE and ALMA profile. Then, for each family of models (F1 to F5), we compute a kernel density estimation (KDE), summing normal distributions with a standard deviation of $1$\,au, centered at the value of the offset. This is done to avoid sensitivity to the bin size and provide a more continuous representation of the data. The results are shown in Figure\,\ref{fig:kde_size}, for all nine families of models, alongside with the native histograms.

For some of the families, there are offsets larger than $15$\,au. This happens when the secondary ring becomes brighter than the birth ring in scattered light, and is further discussed in Section\,\ref{sec:HD131835}. Barring these particular cases, we see that the maximum of the KDEs shifts towards larger offsets as we decrease $q$ and $\beta_\mathrm{scale}$. Decreasing $s_\mathrm{blow}$ seems to be slightly more efficient than decreasing $q$; the KDE for family F2 peaks at larger distance than the one for family F3. Changing both parameters seems to be cumulative and the KDE for family F4 peaks at $\sim 6.5$\,au. Releasing a larger quantity of small dust particles (steeper size distribution or decreasing the blow-out size via $\beta_\mathrm{scale}$) therefore helps increase the offset between the SPHERE and ALMA radial profiles. We remark that increasing the minimum grain size from $0.1$ to $0.5$\,$\mu$m (F5 compared to F4) yields a similar distribution of offsets. The scattering efficiencies $Q_\mathrm{sca}$ for sub-$\mu$m particles ($s < 0.5$\,$\mu$m) are likely too small to have a significant impact on the resulting surface brightness profiles (see e.g., Figure 4 of \citealp{Olofsson2023}).

\subsection{Gas kinetic temperature and mean molecular weight}\label{sec:tgas}

Since the local sound speed $c_\mathrm{s}$ intervenes when estimating the gas-drag force and depends on both the temperature profile and the mean molecular weight, we then vary these two parameters. We computed new models based on family F4 (Table\,\ref{tab:model_families}), which shows the largest offsets and set $T_\mathrm{g, 0} = 300$\,K and $\mu = 28$ (F6) or $T_\mathrm{g, 0} = 40$\,K with $\mu = 2$ (F7).

Figure\,\ref{fig:kde_size} (for families F4, F6, and F7) shows that changing either parameter has little impact on the radial offset between SPHERE and ALMA. As illustrated in Fig.\,\ref{fig:eta}, for a given radius $r$, the stable region where $\eta(r) = \beta(s)$ (Eqn.\,\ref{eqn:eta}) is reached for larger $\beta$ values when increasing $T_\mathrm{g,0}$ or decreasing $\mu$. This means that, assuming they survive long enough, the small particles will not drift as far away from the birth ring. For the larger grains, contributing to the ALMA profiles, the situation is less straightforward. From Figure\,\ref{fig:eta}, the stability curve for family F4 crosses $r = 75$\,au for $\beta \sim 10^{-3}$, which is shifted to $\beta \sim 10^{-2}$ for models of family F6. All particles with $10^{-3} < \beta < 10^{-2}$ should drift outward for models of family F4, while the same grains should remain close to the birth ring for families F6 and F7. While this should help in obtaining a larger offset for families F6 or F7, the assumption that the particles can survive long enough may not hold. Indeed, the birth ring is where the optical depth peaks, hence where the lifetime of any particle is shortest. Overall, this is a multi-dimensional problem, but results of Fig.\,\ref{fig:kde_size} suggest that the kinetic temperature or mean molecular weight have little impact on the offset between SPHERE and ALMA observations, in the regime $r_\mathrm{SPHERE} - r_\mathrm{ALMA} \lesssim 10$\,au. However, when it comes to the formation of secondary rings, Fig.\,\ref{fig:kde_size} shows that increasing $T_\mathrm{g, 0}$ or decreasing $\mu$ results in fewer systems with offsets larger than $\sim 20$\,au (that is, secondary rings). This is in line with the results presented in \citet{Jankovic2026} who found that the distribution of small particles is smoother for lower $\mu$ values, while grains with sizes $10 < s < 30$\,$\mu$m pile-up outside the birth ring and create a detectable secondary ring  for models with larger $\mu$ values (their Figure\,7).

\subsection{Birth ring and gaseous disk radius}\label{sec:radius}

\begin{figure}
  \centering
  \includegraphics[width=1.0\linewidth]{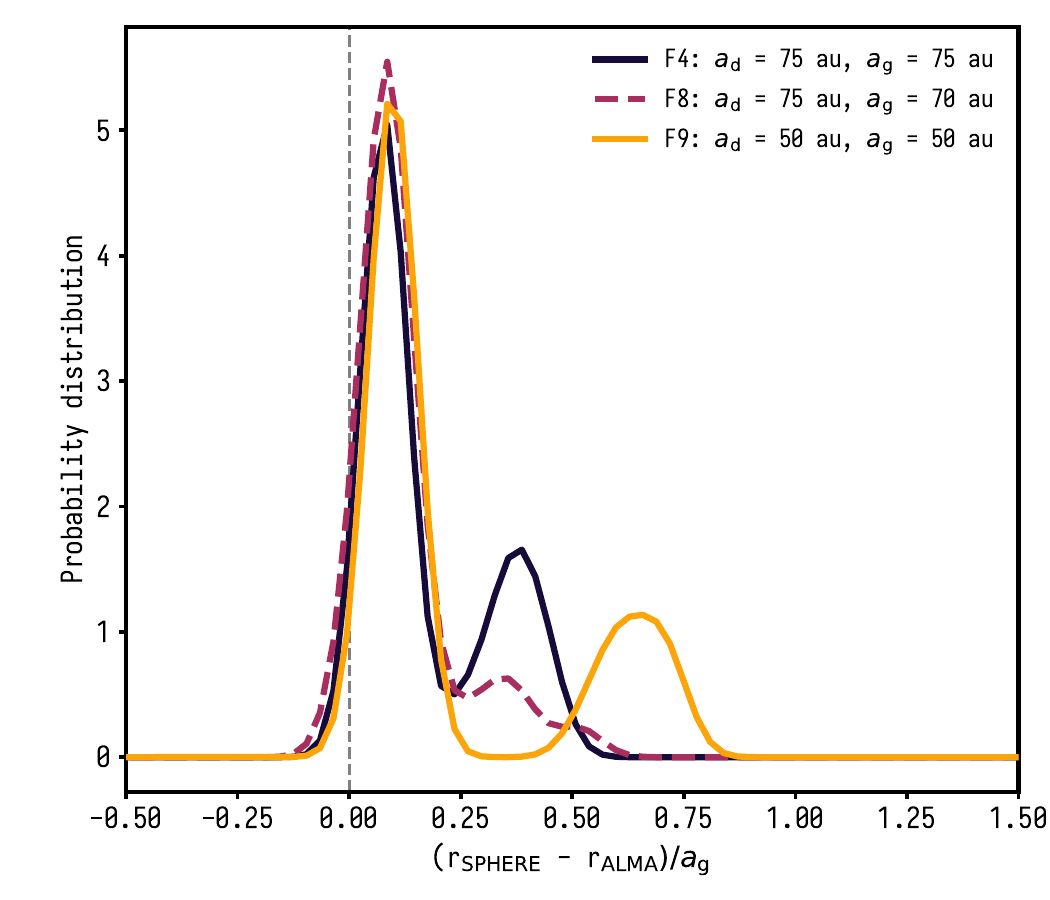}
    \caption{Kernel density estimation of the offset divided by the reference radius of the gaseous disk $a_\mathrm{g}$ for families F4, F8, and F9, using a kernel standard deviation of $0.05$.
  \label{fig:kde_radius}}
\end{figure}

In this subsection we investigate how the scale of the offset varies when changing the reference radius of either the birth ring $a_\mathrm{d}$ or the gaseous disk $a_\mathrm{g}$. Our reference family of models is F4 (Table\,\ref{tab:model_families}), since this family shows the largest offsets (Fig.\,\ref{fig:kde_size}). For family F8, the gaseous disk peaks at $a_\mathrm{g} = 70$\,au instead of $75$\,au, with $a_\mathrm{d} = 75$\,au (we note that for this family of models, $\mu$ is set to $2$, compared to $28$ for F4, but we showed that this only has a small effect, Section\,\ref{sec:tgas}). Family F9 is similar to F4, except that $a_\mathrm{d} = a_\mathrm{g} = 50$\,au. Figure\,\ref{fig:kde_radius} shows the KDEs for these three families of models (with standard deviation of $0.05$ for the kernel), but in this specific case, the offset is divided by $a_\mathrm{g}$ so that we can quantify the relative effect that $a_\mathrm{g}$ has. Even though the extent of the parameter space exploration is limited to a handful of values, our results suggest that the \textit{relative} offset does not depend significantly on $a_\mathrm{g}$ since all KDEs peak at the same location. The offset would be smaller for more compact disks, and larger for disks at a greater separation from the star.

\section{Discussion}\label{sec:discussion}

\subsection{The competition between gas mass and dust optical depth}\label{sec:gmas_tau}

\begin{figure}
  \centering
  \includegraphics[width=1\linewidth]{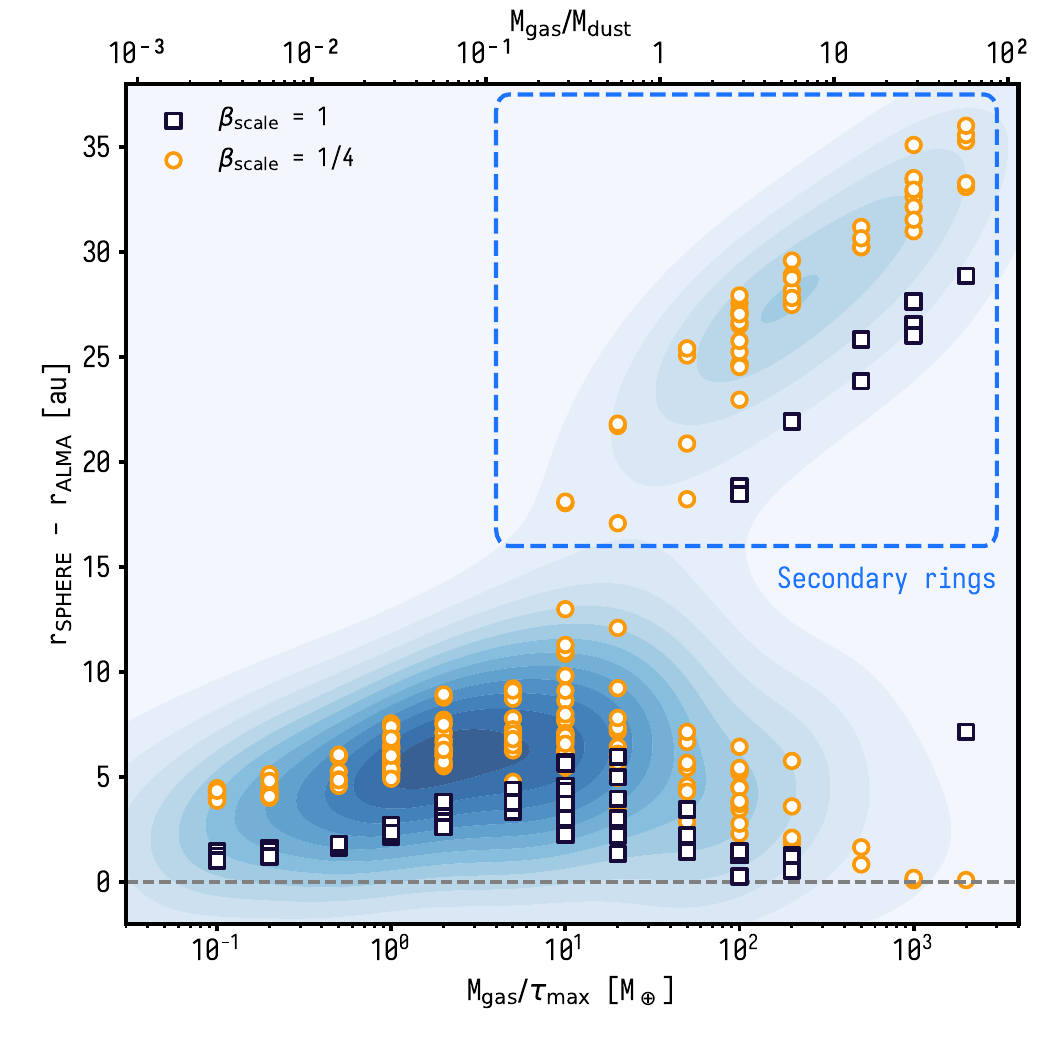}
    \caption{Offset between SPHERE and ALMA as a function of the ratio between the gas mass $M_\mathrm{gas}$ and maximum optical depth $\tau_\mathrm{max}$ for all models from families F1 to F8 (excluding F9). Models with different $\beta_\mathrm{scale}$ are shown with different symbols and colors. The dashed blue box shows the locus of models displaying a secondary belt brighter than the birth ring in the SPHERE radial profiles. The underlying contours show a two-dimensional KDE for all the models. A secondary x-axis at the top shows the estimated gas to dust mass ratio.
  \label{fig:correlation_mgas_tau}}
\end{figure}

Figure\,\ref{fig:correlation_mgas_tau} shows the offset measured between the peak positions of the SPHERE and ALMA radial profiles as a function of the ratio $M_\mathrm{gas}/\tau_\mathrm{max}$. Since $\tau_\mathrm{max}$, which governs the collisional timescale, is directly related to the total dust mass, the ratio $M_\mathrm{gas}/\tau_\mathrm{max}$ can be considered as a proxy for the gas-to-dust mass ratio. We used Equation 7 from \citet{Wyatt2008} to convert the fractional luminosity $f_\mathrm{disk}$ to an estimated dust mass, as
\begin{equation}
    M_\mathrm{dust} = 12.6 f_\mathrm{disk} a_\mathrm{d}^2 \kappa_\nu^{-1} X_\lambda^{-1},
\end{equation}
where $\kappa_\nu$ is the dust opacity ($1.9$\,cm$^2$g$^{-1}$ at $0.89$\,mm, \citealp{Marino2026a}, converted to $50.7$\,au$^2$ $M_\oplus^{-1}$) and $X_\lambda$ is meant to account for the emission falloff at longer wavelengths, which we assumed to be $4$ as in \citet{Wyatt2008}. Since the linear transformation from $\tau_\mathrm{max}$ to $M_\mathrm{dust}$ depends on $a_\mathrm{d}$, we did not include results from family F9, since $a_\mathrm{d}$ is $50$\,au compared to $75$\,au for the rest of the models. The estimated gas to dust mass ratio is shown as the top x-axis on Figure\,\ref{fig:correlation_mgas_tau}. The color-coding and symbols of the scatter plot highlight models with different $\beta_\mathrm{scale}$ values. Both populations display a similar pattern as a function of $M_\mathrm{gas}/\tau_\mathrm{max}$, but with a vertical offset; models with smaller $\beta_\mathrm{scale}$ show overall larger offsets compared to models with $\beta_\mathrm{scale} = 1$, as discussed in Section\,\ref{sec:size}. We also show a two-dimensional KDE underneath the scatter plot, for all the models, to better show the density distribution of offsets.

Initially, as the ratio $M_\mathrm{gas}/\tau_\mathrm{max}$ increases, so does the size of the offset. Larger gas masses will increase the offset as the gas drag force is larger (Equation\,\ref{eqn:gdrag}). Gas drag will both dampen the particles' eccentricities and increase their pericenter distances faster for larger gas masses (e.g., Fig.\,5 in \citealp{Olofsson2022a}). They should thus be less likely to return inside the birth ring where their chances of being destroyed are larger. On the other hand, increasing $\tau_\mathrm{max}$ will decrease the offset, since it influences the collisional lifetimes of the particles (Eqn.\,\ref{eqn:tcoll}). For larger $\tau_\mathrm{max}$ values, particles will be destroyed more efficiently, before they might reach the location where $\beta = \eta$. Lowering the maximum optical depth or increasing the gas mass therefore helps shift the peaks of the surface brightness profiles to larger separations.

\begin{figure}
  \centering
  \includegraphics[width=1\linewidth]{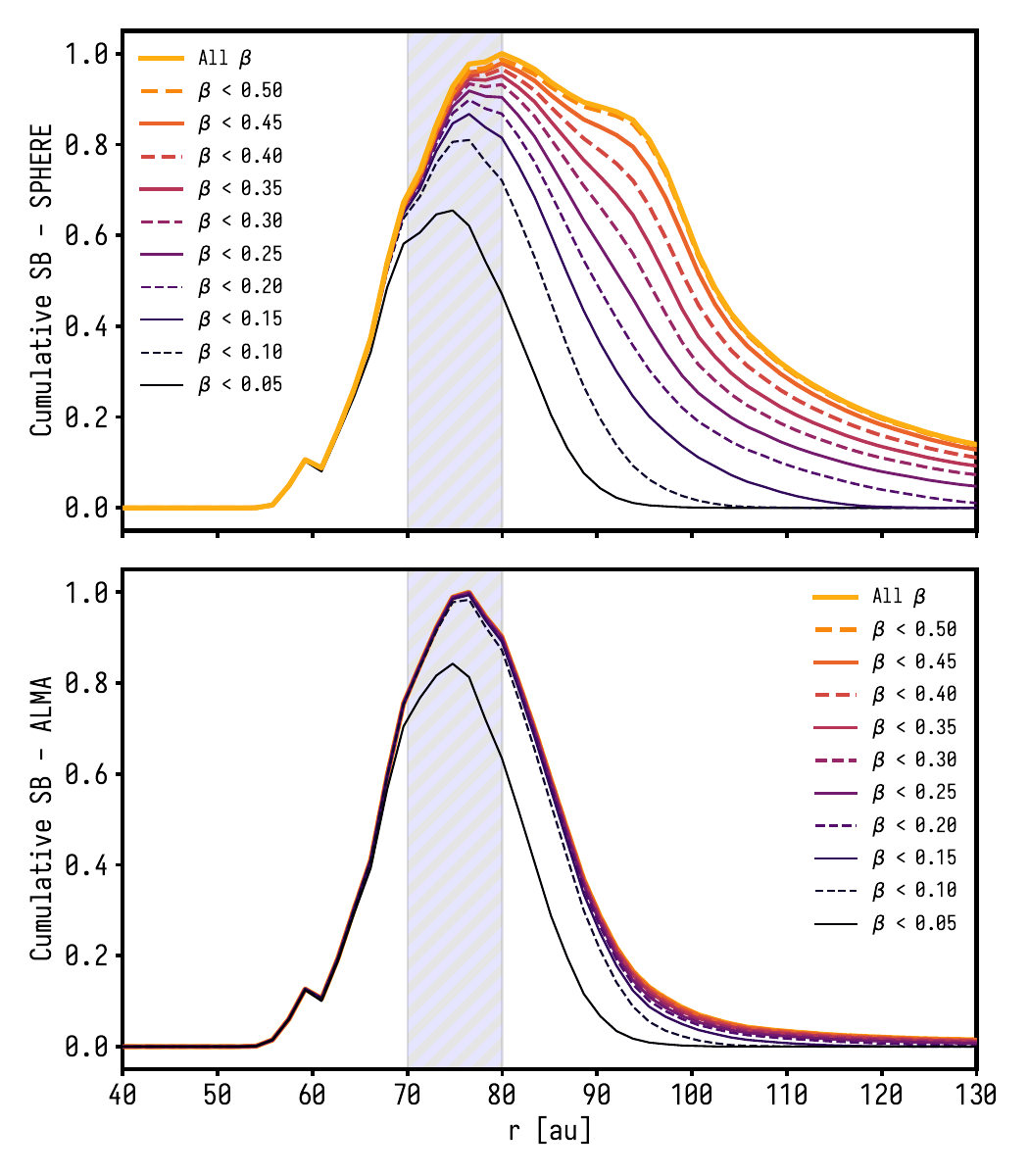}
    \caption{Same as Figure\,\ref{fig:cumulative} for the model of family F1 with $M_\mathrm{gas} = 10^{-2}$\,$M_\oplus$ and $\tau_\mathrm{max} = 5 \times 10^{-4}$.
  \label{fig:shoulder}}
\end{figure}

However, beyond $M_\mathrm{gas}/\tau_\mathrm{max} \sim 10$ ($M_\mathrm{gas}/M_\mathrm{dust} \sim 0.3$) the distribution of offsets splits into two different branches. The upper branch is populated with models that show a secondary ring brighter than the birth ring in scattered light; models with large gas masses or low optical depths where the small dust particles can drift outwards before being destroyed. On the other hand, the lower branch is populated with models where the effect of gas drag results in a broadening of the SPHERE surface brightness profiles, or the creation of a secondary ring which does not become brighter than the birth ring. As an example, the surface brightness profiles for a model with a ``shoulder'' in scattered light are shown in Figure\,\ref{fig:shoulder}. The contribution of the small grains in the SPHERE profile is mostly confined to distances in the range $90-100$\,au. As a consequence, the surface brightness in the birth ring (the hatched region) arises from larger grains, which are less sensitive to gas drag and radiation pressure, resulting in a smaller offset between SPHERE and ALMA.

\subsection{Mid-IR observations}\label{ref:jwst}

\begin{figure*}
  \centering
  \includegraphics[width=1\linewidth]{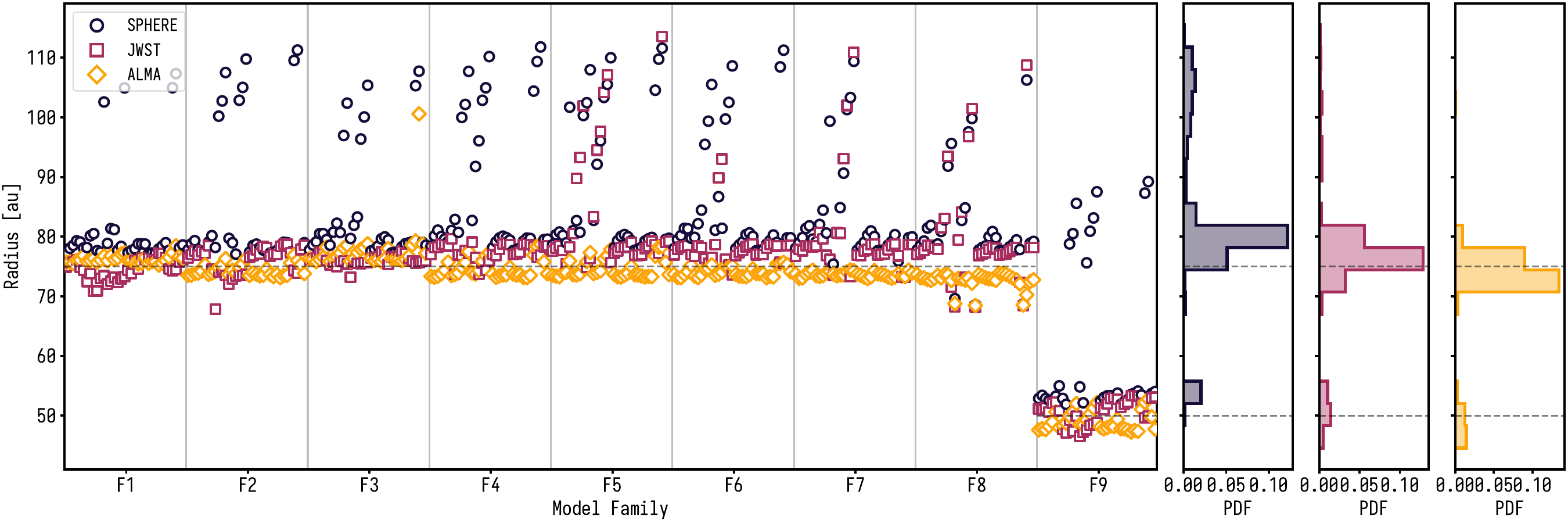}
    \caption{\textit{Left: }Peak positions of the surface brightness for SPHERE, JWST, and ALMA for the different families of models described in Table\,\ref{tab:model_families}. The horizontal dashed line marks $a_\mathrm{d}$ ($75$\,au for families F1 to F8 and $50$\,au for family F9). \textit{Rightmost panels:} histograms of the peak positions for each wavelength regime. 
  \label{fig:jwst}}
\end{figure*}

Mid-IR observations are complementary to mm observations as the thermal emission no longer probes the Rayleigh-Jeans tail and the flux density depends much more on the stellocentric distance. For each model we therefore computed new thermal emission images at a wavelength of $18$\,$\mu$m (akin to JWST/MIRI observations and in the same wavelength regime as ELT/METIS observations). The peak positions of the SPHERE, ALMA, and the newly computed radial profiles (to which we refer to as "JWST") are shown in Figure\,\ref{fig:jwst} arranged by their respective families. The three rightmost panels show the histograms of peak positions for each wavelength.

In most cases the peak positions of the JWST profiles lie in-between the ALMA and SPHERE ones, but there are some exceptions. Interestingly, we see that the JWST profiles peak inward of the ALMA profiles for family F1. This can be explained by computing the flux density at $18$ and $880$\,$\mu$m, which is shown in Figure\,\ref{fig:planck}, for a given grain size ($100$\,$\mu$m). The flux density decreases much faster with the stellocentric distance in the mid-IR compared to sub-mm. For the remaining families of models, this effect makes a smaller difference. The different behavior of family F1 compared to all the other models is related to the abundance of small dust grains. Family F1 is the only one for which $q = -3.5$ and $\beta_\mathrm{scale} = 1$, and it is the only family for which the JWST profiles mostly peak inward of the ALMA profiles.

We can also notice in Figure\,\ref{fig:jwst} that secondary rings can become brighter than the birth ring at mid-IR wavelengths, for families F6 to F8. The common denominator between these families are the gas properties, either the temperature $T_0$ or the mean molecular weight $\mu$. The brightness ratio between any secondary ring and the main ring at mid-IR wavelength could therefore provide additional constraints on the origin of the gas, informing whether it is of primordial or secondary origin.

\begin{figure}
  \centering
  \includegraphics[width=1\linewidth]{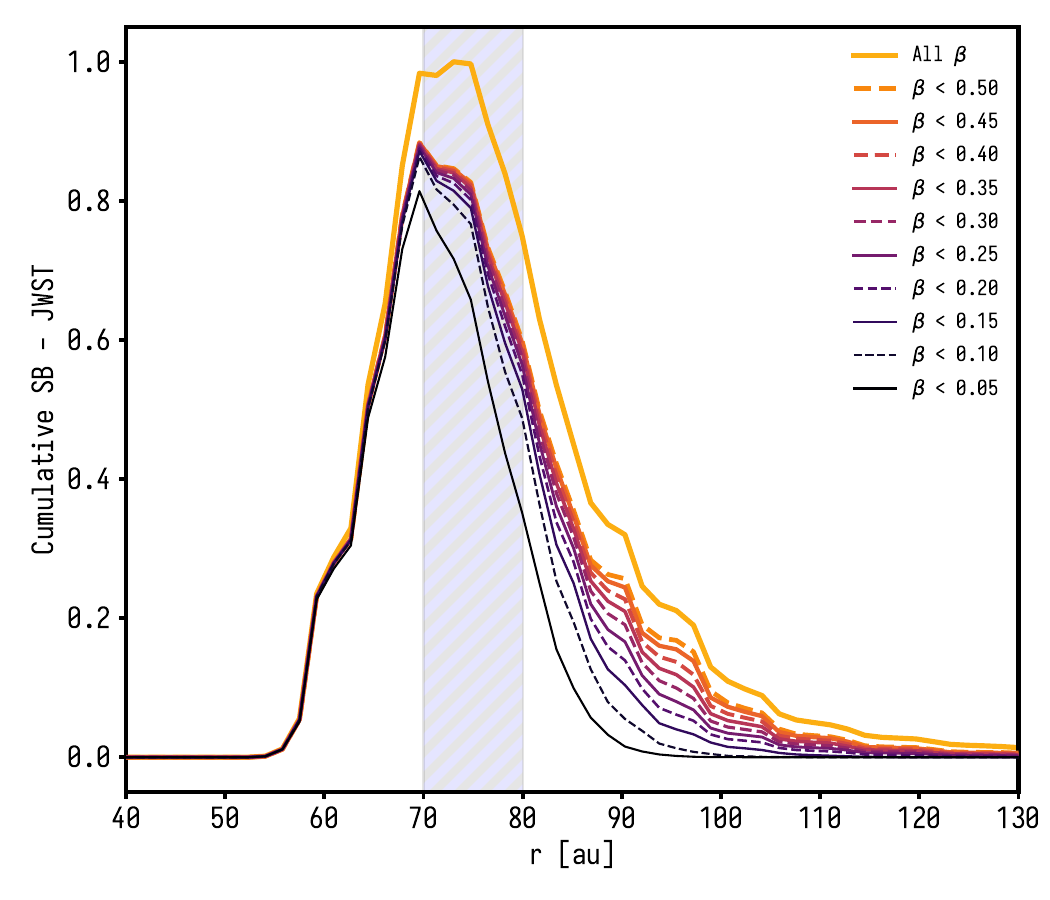}
    \caption{Same as Figure\,\ref{fig:cumulative}, for the thermal emission surface brightness profile at $18$\,$\mu$m, for a model of family F1 ($M_\mathrm{gas} = 0.5$\,$M_\oplus$, $\tau_\mathrm{max} = 10^{-3}$).
  \label{fig:cumulative_jwst}}
\end{figure}

To better understand why the mid-IR peak positions seem to depend on the size distribution (e.g., families F1 vs F4), Figure\,\ref{fig:cumulative_jwst} shows the cumulative surface brightness profiles at $18$\,$\mu$m. Similarly to Fig.\,\ref{fig:cumulative}, the contributions of the different intervals of $\beta$ are shown. For this example, most of the surface brightness can be explained by the contributions of either large grains ($\beta < 0.05$) or much smaller, unbound grains. As we vary either $\beta_\mathrm{scale}$ or $q$ (or both, family F4) to have a larger number of small dust particles, the contribution of unbound grains will increase, shifting the peak to larger separations. This is in line with the results presented in \citet{Thebault2019}, showing that the unbound grains can significantly contribute to the mid-IR flux. This underlines the importance of obtaining high angular resolution observations at these wavelengths, as they probe a population of dust particles that are otherwise poorly constrained.

Any further analysis of the mid-IR profiles would require the implementation of Poynting-Robertson drag in the code. Recent JWST observations of debris disks such as Vega (\citealp{Su2024}) and Fomalhaut (\citealp{Sommer2025}) suggest that PR drag plays a significant role in shaping the morphology of the disks as seen at these wavelengths (even though these two particular systems have smaller fractional luminosities than the models discussed in this study).

\subsection{Secondary rings}\label{sec:HD131835}

\begin{figure*}
  \centering
  \includegraphics[width=1\linewidth]{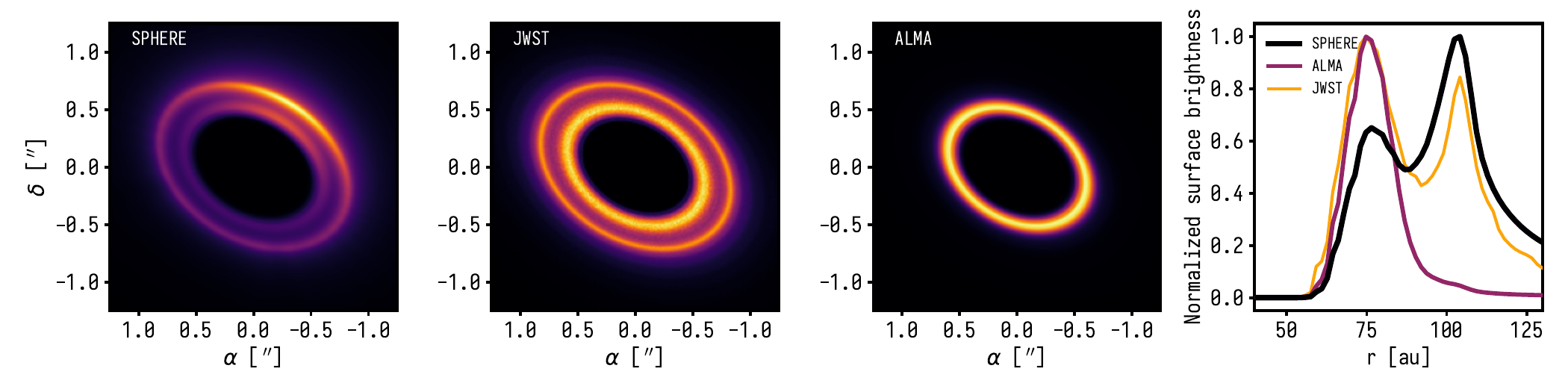}
    \caption{\textit{Left to right:} SPHERE, JWST, and ALMA images, as well as surface brightness profiles for all three wavelengths, for a model from family F4, with a gas mass of $5 \times 10^{-2}$\,$M_\oplus$ and a maximum optical depth of $5 \times 10^{-4}$. The pixel size is $12.26$\,milli-arcseconds for all three images (akin to SPHERE/IRDIS) and no convolution by a point spread function has been applied.
  \label{fig:HD131835}}
\end{figure*}

\citet{Milli2026} compared the radial profiles derived from SPHERE and ALMA observations for the ARKS sample and showed that the disk around HD\,131835 is a rather unique system. Among the six disks that display a significant offset between scattered light and sub-mm observations, HD\,131835 has the largest one: the ALMA profile peaks at $65$\,au (from the \texttt{frank} modeling results) to be compared to $112$\,au in scattered light for the brightest ring, leading to a relative offset of $\sim 70$\%. \citet{Jankovic2026} investigated different scenarios to explain the origin of this large offset. One of them is the effect of gas-drag, on which we based our calculations of $\eta(r)$. However, there are some differences between the two approaches: \citet{Jankovic2026} do not directly account for collisions nor for the presence of unbound grains, while we do not include the effect of PR drag in our simulations.

Some of our models show first order similarities with the observations of HD\,131835. Figure\,\ref{fig:HD131835} shows one of the models from family F4, with a gas mass of $5 \times 10^{-2}$\,$M_\oplus$ and a maximum optical depth of $\tau_\mathrm{max} = 5 \times 10^{-4}$. This combination of parameters ensures that the grains can survive for a long time while experiencing non-negligible gas drag. The SPHERE, JWST, and ALMA images are displayed, from left to right, while the rightmost panel shows their corresponding surface brightness profiles. At mm wavelengths, only the main ring is visible, even though from the surface brightness there might be a hint of a secondary bump close to $100-110$\,au. At mid-IR wavelengths, the main ring appears very similar to the one seen with ALMA, but the secondary ring is almost as bright ($\sim 70$\%) in thermal emission and is clearly visible in the image. In scattered light, the secondary ring is brighter than the main ring, echoing what is observed for HD\,131835 (\citealp{Feldt2017}, \citealp{Milli2026}, \citealp{Jankovic2026}). Interestingly, we can see in the image that the azimuthal brightness due to the scattering phase function is different in the inner and outer rings. This is because the inner ring mostly consists of large grains ($\beta < 0.05$) while the scattered light from the outer ring arises from particles with $\beta > 0.05$. 

We note that HD\,131835 is not the only debris disk for which a secondary ring has been reported from near-IR scattered light observations. \citet{Olofsson2023} reported the detection of a faint outer ring in the disk around HD\,129590. This ring is only detected in total intensity, not in polarized light, and the disk is known to harbor cold CO gas (\citealp{Kral2020}). Similar results were found for HD\,120326, with the exception that the disk seems to be devoid of cold CO gas (\citealp{Bonnefoy2017}, \citealp{Desgrange2025}). While it is often considered as a ``hybrid'' disk, the disk around HD\,141569 is known to be gas-rich (\citealp{Flaherty2016}) while showing multiple rings in scattered light (\citealp{Singh2021}). However, none of these systems currently have high angular resolution observations at mm wavelengths that can be compared with the scattered light observations.

Interestingly, when the gas mass is large and the optical depth small, we find that some models also exhibit a secondary ring in the ALMA surface brightness profiles (though they never become brighter than the birth ring). Regardless of the wavelength at which a secondary ring is observed, such bimodal radial profiles could be misinterpreted as the presence of a gap in the distribution of \textit{planetesimals}, mimicking the presence of massive perturbers in the disk. That being said, in a second generation gas scenario, it is unclear how we could have, at the same time, low optical depth and large gas mass since the gas should be released from collisions of icy bodies which would also produce significant amounts of dust (e.g., \citealp{Marino2020}).

These results should however be taken with caution since our simulations do not include the effect of PR drag. As mentioned in \citet{Jankovic2026}, PR drag starts becoming important in regions where the gas density is low (but has little impact in regions of higher density). This is best illustrated in their Figure\,7, showing grains can stop drifting outwards before they reach the regions where $\beta (r) = \eta (r)$. In our simulations drift stops either at $\beta (r) = \eta (r)$ or where the gas density and drift velocity drop to very low values. In practice, this suggests that the secondary ring could be closer to the birth ring compared to our results. Future works focusing on the properties of these secondary rings should include PR drag (as in \citealp{Jankovic2026}) and collisions (as in this study).

\section{Summary}\label{sec:summary}

With new high angular resolution observations we can probe the spatial distribution of dust particles at unprecedented details, opening new ways to study debris disks by following the dynamics of the dust grains. Motivated by the comparison between new ARKS sub-mm and archival SPHERE near-IR observations of gas-bearing debris disks (\citealp{Milli2026}), we investigated how the distribution of dust particles is affected by gas drag. The metric of interest in this study is the radial offset between peak surface brightness in images at different wavelengths. We presented a new parameter space study, building upon past works (\citealp{Takeuchi2001}, \citealp{Krivov2009}, \citealp{Olofsson2022a}, \citealp{Jankovic2026}) and we summarize our main results here.

\begin{itemize}
\item Gas drag can affect the dynamics of dust particles, and the final dust density distribution strongly depends on both the gas mass and the disk's optical depth. Increasing the gas mass increases the rate at which particles drift outwards, while a larger optical depth implies shorter collisional lifetimes, limiting how far particles can drift. 
\item Even though we investigated a small part of the parameter space, we find that variations in the gas mean molecular weight and kinetic temperature do not appear to have a significant impact on the extent of the offset.
\item We find that either having a steeper size distribution ($q < -3.5$) or decreasing the blow-out size can increase the radial offset measured from multi-wavelength observations, compatible with the offsets reported in \citet{Milli2026}.
\item We show that mid-IR wavelengths are complementary to near-IR and sub-mm observations and measuring radial offsets between the three different combinations could provide additional constraints on the gaseous disk's properties and the grain size distribution.
\item While the contribution of unbound grains is negligible at sub-mm wavelengths, they do contribute in near-IR scattered light and even more so at mid-IR wavelengths, in thermal emission. Their contribution should not be ignored in this wavelength regime.
\item When no secondary ring is observed, we find that the extent of the offset scales with the reference radius of the disk: smaller disks should display smaller offsets.
\item Similarly to previous work we find that a secondary ring can appear outwards from the planetesimal belt. This only occurs under the right conditions (low optical depth and high gas mass) and the ring appears in both scattered light and mid-IR observations (and be brighter than the birth ring), while it remains barely detectable at mm wavelengths. Models with low mean molecular weight or large kinetic temperature seem to be less prone to developing a secondary ring.
\end{itemize}

The approach described in this work is currently not well suited to directly fit observations, due to the relatively high computational time required, as well as regions of the parameter space that we have not explored (e.g., dust composition). That being said, the observations obtained with the ARKS Large Program, combined with near-IR scattered light observations, opened a new avenue to study the dynamics of dust particles, and this new diagnostic can help us better understand how dust particles evolve after being produced from collisions of planetesimals in debris disks.

\begin{acknowledgements}
We thank the referee for their thorough report that helped improving the presentation of our results, especially for the comparison between the different families of models.
    This research made use of Astropy (\url{http://www.astropy.org}) a community-developed core Python package for Astronomy \citep{astropy:2013, astropy:2018}, Numpy (\citealp{numpy}), Matplotlib (\citealp{matplotlib}), and Numba (\citealp{numba}).
This paper makes use of the following ALMA data: ADS/JAO.ALMA\# 2022.1.00338.L, 2012.1.00142.S, 2012.1.00198.S, 2015.1.01260.S, 2016.1.00104.S, 2016.1.00195.S, 2016.1.00907.S, 2017.1.00167.S, 2017.1.00825.S, 2018.1.01222.S and 2019.1.00189.S. ALMA is a partnership of ESO (representing its member states), NSF (USA) and NINS (Japan), together with NRC (Canada), MOST and ASIAA (Taiwan), and KASI (Republic of Korea), in cooperation with the Republic of Chile. The Joint ALMA Observatory is operated by ESO, AUI/NRAO and NAOJ. The National Radio Astronomy Observatory is a facility of the National Science Foundation operated under cooperative agreement by Associated Universities, Inc. The project leading to this publication has received support from ORP, that is funded by the European Union’s Horizon 2020 research and innovation programme under grant agreement No 101004719 [ORP]. We are grateful for the help of the UK node of the European ARC in answering our questions and producing calibrated measurement sets. This research used the Canadian Advanced Network For Astronomy Research (CANFAR) operated in partnership by the Canadian Astronomy Data Centre and The Digital Research Alliance of Canada with support from the National Research Council of Canada the Canadian Space Agency, CANARIE and the Canadian Foundation for Innovation.
MRJ acknowledges funding provided by the Institute of Physics Belgrade, through the grant by the Ministry of Science, Technological Development, and Innovations of the Republic of Serbia. SM acknowledges funding by the Royal Society through a Royal Society University Research Fellowship (URF-R1-221669) and the European Union through the FEED ERC project (grant number 101162711). MB acknowledges funding from the Agence Nationale de la Recherche through the DDISK project (grant No. ANR-21-CE31-0015). AMH acknowledges support from the National Science Foundation under Grant No. AST-2307920. SMM acknowledges funding by the European Union through the E-BEANS ERC project (grant number 100117693), and by the Irish research Council (IRC) under grant number IRCLA- 2022-3788. Views and opinions expressed are however those of the author(s) only and do not necessarily reflect those of the European Union or the European Research Council Executive Agency. Neither the European Union nor the granting authority can be held responsible for them. EM acknowledges support from the NASA CT Space Grant. LM acknowledges funding by the European Union through the E-BEANS ERC project (grant number 100117693), and by the Irish research Council (IRC) under grant number IRCLA- 2022-3788. Views and opinions expressed are however those of the author(s) only and do not necessarily reflect those of the European Union or the European Research Council Executive Agency. Neither the European Union nor the granting authority can be held responsible for them. JM acknowledges funding from the Agence Nationale de la Recherche through the DDISK project (grant No. ANR-21-CE31-0015) and from the PNP (French National Planetology Program) through the EPOPEE project. A.A.S. is supported by the Heising-Simons Foundation through a 51 Pegasi b Fellowship. Support for BZ was provided by The Brinson Foundation. CdB acknowledges support from the Spanish Ministerio de Ciencia, Innovaci\'on y Universidades (MICIU) and the European Regional Development Fund (ERDF) under reference PID2023-153342NB-I00/10.13039/501100011033, from the Beatriz Galindo Senior Fellowship BG22/00166 funded by the MICIU, and the support from the Universidad de La Laguna (ULL) and the Consejer\'ia de Econom\'ia, Conocimiento y Empleo of the Gobierno de Canarias. JBL acknowledges the Smithsonian Institute for funding via a Submillimeter Array (SMA) Fellowship, and the North American ALMA Science Center (NAASC) for funding via an ALMA Ambassadorship. TDP is supported by a UKRI Stephen Hawking Fellowship and a Warwick Prize Fellowship, the latter made possible by a generous philanthropic donation. SP acknowledges support from FONDECYT Regular 1231663 and ANID -- Millennium Science Initiative Program -- Center Code NCN2024\_001.

\end{acknowledgements}

\bibliographystyle{aa}
\bibliography{biblio}

\appendix

\section{Dust temperature, $\beta$, and $\eta$}\label{app:model}

\begin{figure}
  \centering
  \includegraphics[width=1.0\linewidth]{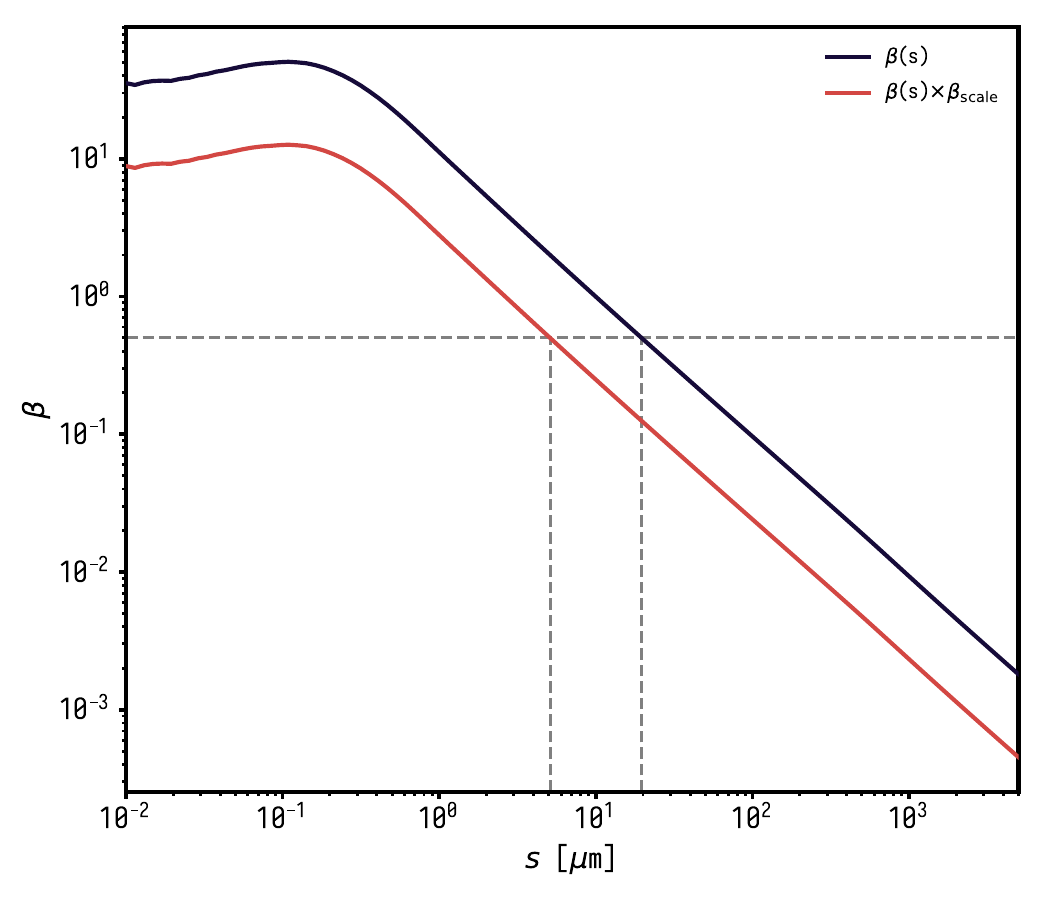}
    \caption{Radiation pressure strength $\beta$(s) for the default composition and stellar parameters (black solid line). Scaling down the original curve by $\beta_\mathrm{scale}$ effectively changes the blow-out size $s_\mathrm{blow}$ (red curve). The horizontal dashed line shows $\beta = 0.5$, while the vertical dashed lines show the blow-out size for both $\beta (s)$.
  \label{fig:beta}}
\end{figure}

\begin{figure}
  \centering
  \includegraphics[width=1.0\linewidth]{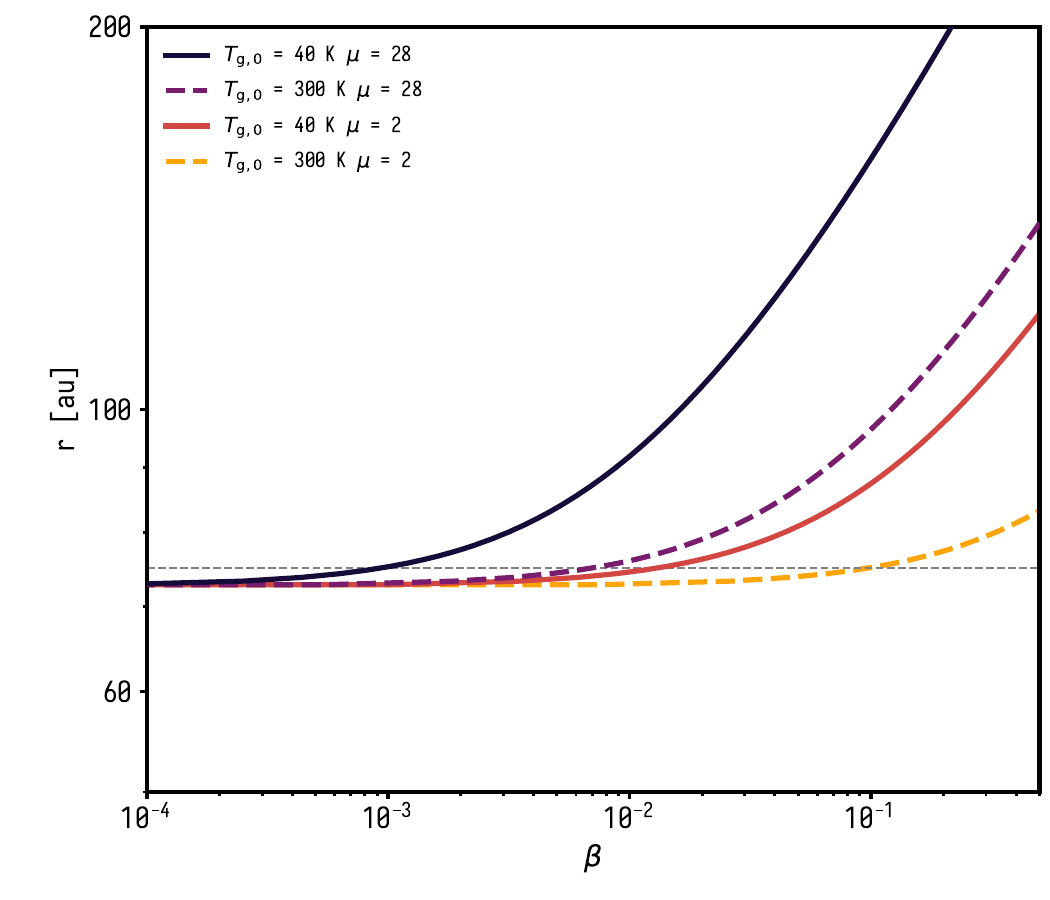}
    \caption{Stability curves in the $\beta(r)$ plane, for different configurations of the reference gas temperature and mean molecular weight. A particle released at a given $(r, \beta)$ above (below) one of the curves will drift inwards (outwards), until it reaches the curve. The horizontal dashed line represents $a_\mathrm{d}$, the location of the birth ring.
  \label{fig:eta}}
\end{figure}

\begin{figure}
  \centering
  \includegraphics[width=1.0\linewidth]{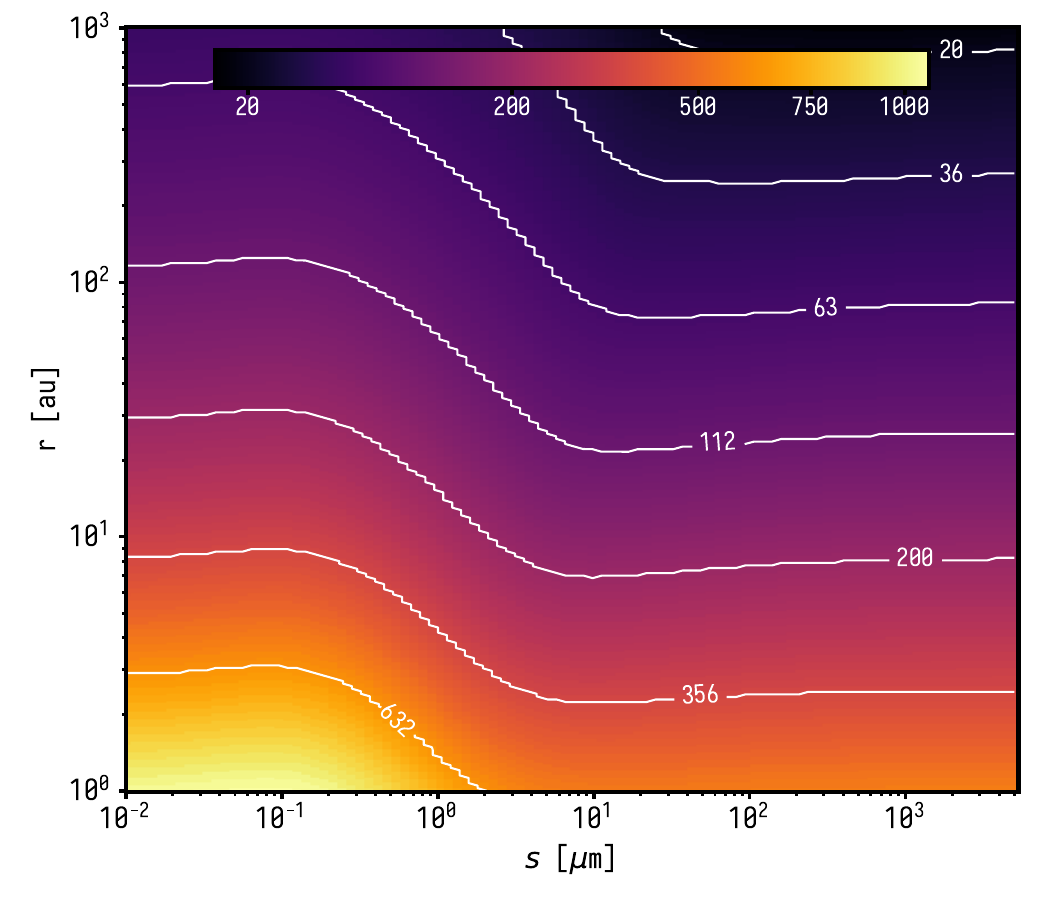}
    \caption{Example map of $T_\mathrm{dust}$ as a function of the grain size and stellocentric distance, with iso-temperature contours.
  \label{fig:tdust}}
\end{figure}

\begin{figure}
  \centering
  \includegraphics[width=1.0\linewidth]{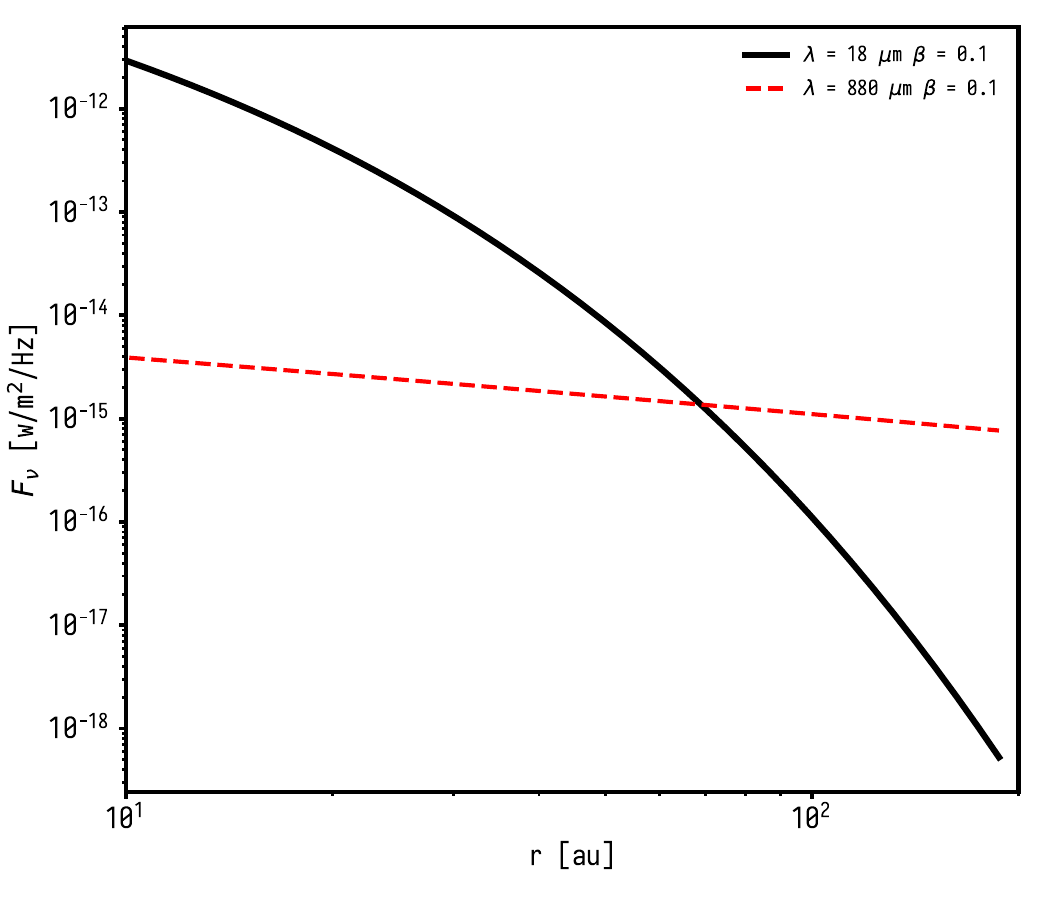}
    \caption{Flux density as a function of stellocentric distance, for a particle of size $s = 100$\,$\mu$m ($\beta \sim 0.1$) for two different wavelengths, $18$ and $880$\,$\mu$m.
  \label{fig:planck}}
\end{figure}

We here show additional diagnostic plots providing context for the modeling approach. The $\beta(s)$ and $\eta(r)$ curves are shown in Figs.\,\ref{fig:beta} and \ref{fig:eta}, respectively. Figure\,\ref{fig:tdust} shows how the temperature of the dust particles varies as a function of the stellocentric distance $r$ and grain size $s$ and Figure\,\ref{fig:planck} shows how the thermal emission flux density varies as a function of $r$ for a specific grain size ($s = 100$\,$\mu$m) at two different wavelengths ($18$ and $880$\,$\mu$m, respectively).

\end{document}